\documentclass[preprint2]{aastex62} 

\graphicspath{{./}{Figures/}}
\revised{\today}
\shorttitle{Limit on Infared Emission from the Galactic Black Hole Accretion Flow}
\shortauthors{Ciurlo et al.}

\begin{document}

\title{Upper Limit on Brackett-$\gamma$ Emission from the Immediate Accretion Flow onto the Galactic Black Hole}
\correspondingauthor{Anna Ciurlo}
\email{ciurlo@astro.ucla.edu}
\author[0000-0001-5800-3093]{Anna Ciurlo}
\affil{Department of Physics and Astronomy, University of California, Los Angeles, CA 90095, USA}
\author[0000-0002-6753-2066]{Mark R.\ Morris}  
\affiliation{Department of Physics and Astronomy, University of California, Los Angeles, CA 90095, USA}
\author[0000-0002-3289-5203]{Randall D. Campbell}
\affiliation{W. M. Keck Observatory, Waimea, HI 96743, US}
\author[0000-0003-3230-5055]{Andrea M. Ghez}
\affiliation{Department of Physics and Astronomy, University of California, Los Angeles, CA 90095, USA}
\author[0000-0001-9554-6062]{Tuan Do}
\affiliation{Department of Physics and Astronomy, University of California, Los Angeles, CA 90095, USA}
\author[0000-0003-3765-8001]{Devin S. Chu}
\affiliation{Department of Physics and Astronomy, University of California, Los Angeles, CA 90095, USA}

\begin{abstract}
We present the first observational constraint on the Br$\gamma$ recombination line emission associated with the supermassive black hole at the center of our Galaxy, known as Sgr~A*.     
By combining 13 years of data with the Adaptive Optics fed integral field  spectrograph OSIRIS at the W. M. Keck Observatory obtained as part of the Galactic Center Orbits Initiative, we extract the near-infrared spectrum within $\sim$0.”2 of the black hole and we derive an upper limit on the Br$\gamma$ flux.
The aperture was set to match the size of the disk-like structure that was recently reported based on millimeter-wave ALMA observations of the hydrogen recombination line, H30$\alpha$.   
Our stringent upper limit is at least a factor of 80 (and up to a factor of 245) below what would be expected from the ALMA measurements and strongly constrains possible interpretation of emission from this highly under-luminous supermassive black hole.
\end{abstract}



\section{Introduction} \label{sec:intro}

The center of our Galaxy hosts a supermassive black hole (SMBH) whose position coincides with the radio and infrared source, Sgr~A* \citep{Ghez00, Schodel02}.
Thanks to its proximity, this environment presents a unique opportunity to study in detail the accretion mechanism onto a SMBH.
The SMBH is not very active ($10^{-7} L_{Edd}$, \citealt{Baganoff03}) even though it was likely much more active at various times in the past few hundred years \citep{Clavel13, Terrier18} and a recent unusual peak in its activity has been reported in the infrared \citep{Do19}.
The accretion rate of hot gas through the Bondi radius ($\sim$1'') of the central SMBH can be estimated from X-ray observations \citep{Baganoff03}, but the amount of cooler gas very near the black hole is not strongly constrained. 

The low activity of Sgr~A* (e.g. \citealt{Witzel12}) places it in the regime of radiatively inefficient accretion flows (RIAF).
Most models of the accretion flow predict or allow for the formation of a disk (see \citealt{Yaun14} for a review).
Recent modeling of the accretion flow by \cite{Calderon20}, assuming that the accreted material comes from the winds of nearby young, windy stars, predicts a disk phase in the accretion flow.
However, the theoretical formation of a disk is not universally agreed upon.
\cite{Ressler20} also modeled the accretion flow supplied by stellar winds from nearby Wolf-Rayet stars but they find that a disk would not form around Sgr~A*.
In their modeling, they argue that the magnetic field drives a polar outflow that removes angular momentum, and that the fast inflow and outflow of the gas, together with inefficient cooling, prevent the material from circularizing. 

Emission from the accretion flow onto Sgr~A* has recently been detected with ALMA observations by \cite{Murchikova19}, who reports a very broad ($>$2000 km s$^{-1}$) emission feature in the H30$\alpha$ radio recombination line arising within $\sim$0.2--0.3" of Sgr~A*.  
The presence of a velocity gradient across the marginally resolved source led them to conclude that they had observed a disk that is approximately edge-on. Given the reported observation of H30$\alpha$ ($n$ = 31$\rightarrow$30 transition), one would expect significant emission in the lower-lying Br$\gamma$ recombination line (7$\rightarrow$4 transition) at 2.1661 $\mu$m.
 
In this paper, we search for the expected Br$\gamma$ emission based on data gathered by Galactic Center Orbits Initiative.
In Section~\ref{sec:data}, we briefly describe the data used.
Section~\ref{sec:analysis}, describes the analysis procedure to extract the Sgr~A* NIR spectrum. 
In Section~\ref{sec:res} we report our flux limit. 
A preliminary version of our limit on Br$\gamma$ was used by \cite{Murchikova19} to argue that the H30$\alpha$ emission must be augmented by weak maser action. 
In Section~\ref{sec:discuss}, we discuss our limit in comparison to the reported H30$\alpha$ flux and to accretion flow theories. Our conclusions are reported in Section~\ref{sec:concl}.


\section{Dataset} \label{sec:data}
The data for this study were taken as part of the Galactic Center Orbits Initiative (GCOI; PI Ghez, W. M. Keck Observatory 1995 - present). 
We used a subset of the deep observations taken between 2006 and 2018 with OSIRIS, the integral field spectrograph fed by a laser guide star adaptive optics system at W. M. Keck Observatory \citep{Larkin06, Wizinowich06}.  
All included epochs consist of 900-second exposures in Kn3 band (2.121--2.229$~\mu$m) with a 35 mas platescale.
In this setup, the average spectral resolution is R=3800, and at the wavelength of Br$\gamma$ this corresponds to 6~10$^{-4}$~$\mu$m, or 2.5~channels.   

The initial stages of data analysis (including sky subtraction and flux calibration) 
have been described in previous publications \citep{Ghez08, Do09, Do19GR, Ciurlo20}. 
We begin with the flux- and astrometrically- calibrated, mosaiced data-cubes. 
In this work, we only examined a sub-set of GCOI raw data that had already been analyzed with a procedure optimized for gas detection reported in \cite{Ciurlo20}, which differs from that used for the measurements of stellar radial velocities\footnote{While GCOI work on stellar spectra have been extracted from individual dark-subtracted exposures, the spectra in this study are extracted from sky-subtracted mosaics (individual exposures are sky-subtracted before building the mosaic). Using the mosaics allows us the sky coverage necessary to measure the positions of reference stars that we then use to put all observed fields in the same reference frame (for more details see Appendix~\ref{App:cont} and \citealt{Ciurlo20}).}.
Furthermore, we exclude epochs showing strong OH line subtraction residuals (visible as extremely strong absorption lines), which leaves a total of 19 epochs that are summarized in Table~\ref{tab:obs}.  
These measurements have an average spatial resolution of 73~mas, based on a Gaussian fit to S0-2 in the mosaicked data-cubes. This subset of data is optimized for extracting the spectrum of the region around Sgr~A*.

\begin{table}[h]
    \centering
    \begin{tabular}{|c|c|c|c|c|}
    \hline
Date\footnote{This is a subset the GCOI dataset (e.g. \citealt{Do19GR}) as described in Section~\ref{sec:data}.}  & Frames &	FWHM & Aperture & Noise \\
& number & [mas] & fraction & [mJy] \\
\hline
2006-06-18$^{a}$	&	9	&	81 	&  97\% & 0.049 \\ 
2006-06-30$^{a}$	&	9	&	77 	&  82\% & 0.061 \\ 
2008-07-25$^{b}$	&	11	&	81 	&  72\% & 0.041 \\ 
2009-05-05$^{b}$	&	12	&	70 	&  56\% & 0.046 \\ 
2009-05-06$^{b}$	&	12	&	81 	&  71\% & 0.047 \\ 
2010-05-05$^{b}$	&	6	&	70 	&  60\% & 0.060 \\ 
2010-05-08$^{b}$	&	11	&	79 	&  55\% & 0.093 \\ 
2011-07-10$^{b}$	&	6	&	71 	&  47\% & 0.059 \\ 
2012-07-22$^{b}$	&	9	&	92 	&  71\% & 0.064 \\ 
2017-05-17$^{c}$	&	11	&	73 	&  67\% & 0.062 \\ 
2017-05-19$^{c}$	&	6	&	86 	&  62\% & 0.059 \\ 
2017-07-19$^{c}$	&	12	&	77 	&  69\% & 0.067 \\ 
2017-07-27$^{c}$	&	13	&	89 	&  76\% & 0.043 \\ 
2017-08-14$^{c}$	&	8	&	75 	&  67\% & 0.048 \\ 
2018-04-24$^{c}$	&	7	&	73 	&  57\% & 0.068 \\ 
2018-05-23$^{c}$	&	14	&	91 	&  55\% & 0.050 \\ 
2018-07-22$^{c}$	&	11	&	77 	&  68\% & 0.034 \\ 
2018-07-31$^{c}$	&	11	&	73 	&  66\% & 0.038 \\ 
2018-08-11$^{c}$	&	9	&	79 	&  64\% & 0.034 \\ 
\hline
    \end{tabular}
    \caption{Table of observations obtained with OSIRIS in Kn3 band, 35 mas plate-scale. 
    All integration times are equal to 900 s. 
    The reported aperture is the fraction of the total 0.23'' radius aperture that is not masked by our noise cut. 
    The reported noise is per resolution element and it is calculated on the spectrum extracted after applying the mask but rescaled to the effective area of our aperture (i. e. dividing by the fraction of unmasked pixels).
    References: $a$) \citet{Ghez08}, $b$) \citet{Boehle16}, $c$) \citet{Do19GR}.}
    \label{tab:obs}
\end{table}

\section{Analysis} \label{sec:analysis}
\begin{figure*}
    \centering
    \includegraphics[width=10cm]{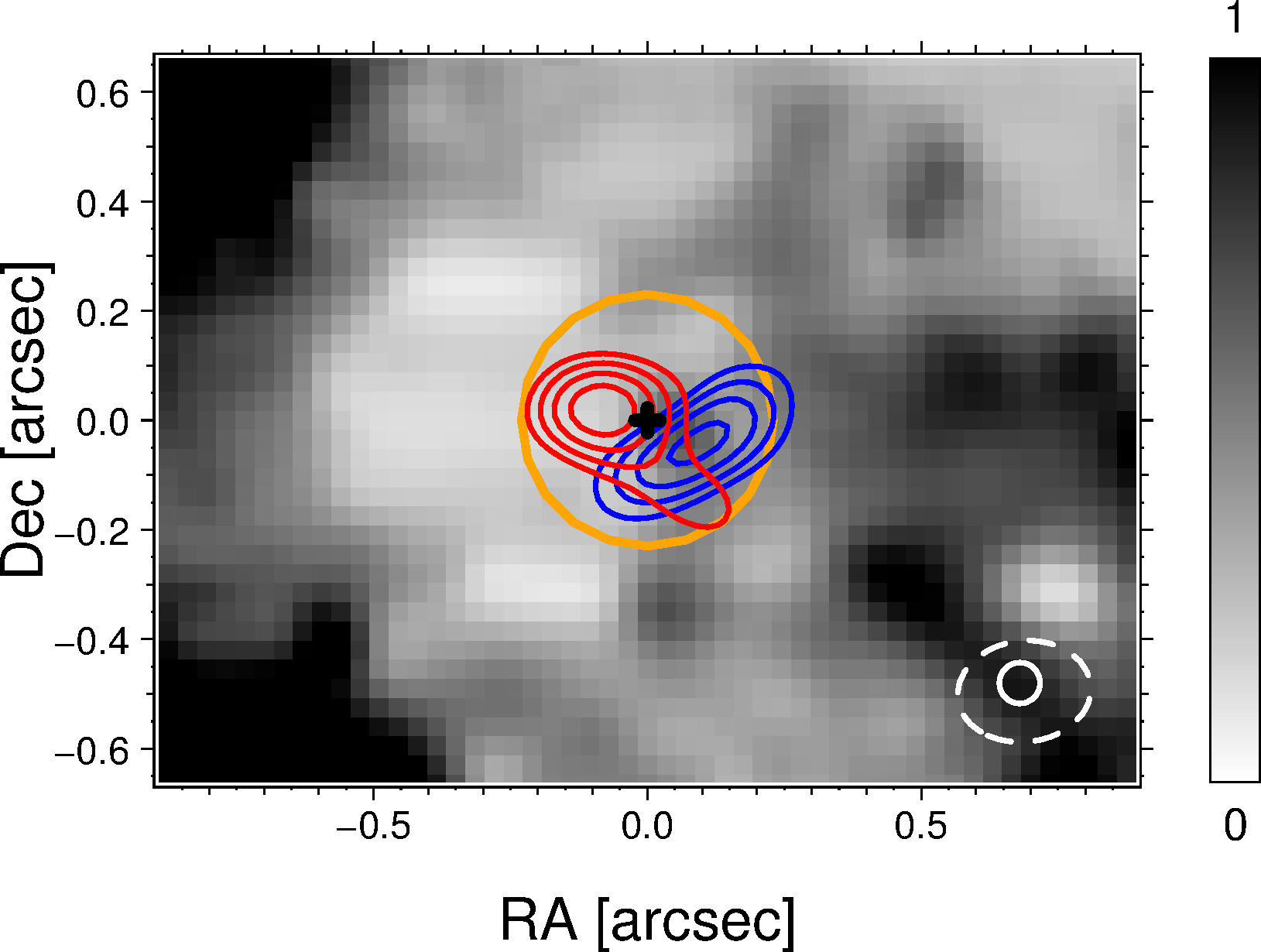}
    \caption{
    Br$\gamma$ emission map in grayscale from the 2017 OSIRIS continuum-subtracted cube, obtained by collapsing the cube between -1100~km/s and 1100~km/s (displayed in arbitrary units).  
    Contours of the reported H30$\alpha$ red and blue-shifted emissions \citep{Murchikova19} are overlaid (the full spectrum is shown in Figure~\ref{fig:2lines}).     
    The blue contours are 1.2, 1.6, 2.0 and 2.4~mJy.     
    The red contours are 0.9, 1.1, 1.3 and 1.5~mJy.  
    The white dashed circle represents the half-power beam for the H30$\alpha$ observations.
    The white solid circle represents the average FWHM of the NIR observations. 
    The position of Sgr~A* is marked by the black cross. The aperture used for this analysis is shown in orange.     
    }
    \label{fig:cont}
\end{figure*}
Our goal is to extract a high signal-to-noise spectrum of the immediate  surroundings of Sgr~A*.
In particular, we seek to establish whether there is a counterpart to the disk-like emission of the reported H30$\alpha$ recombination line.
We therefore extract a spectrum within a circular aperture centered on the position of Sgr~A* and of radius 0.23'' (the same aperture used by \citealt{Murchikova19}), corresponding to a diameter of about 30 typical spatial resolution elements.  
Figure~\ref{fig:cont} shows the chosen aperture along with contours of the H30$\alpha$ emission lines over a map of the Br$\gamma$ emission in the region.
The gas has complex, multi-component emission extended in space and velocity.
We can identify one component to the West, associated with radio source Epsilon \citep{Yusef-Zadeh90}, two in the northeast and southeast corners associated with bright Wolf-Rayet stars (IRS16~C and IRS16~SW) and one featuring an almost linear feature immediately adjacent to the position of Sgr~A* along the line of sight \citep{Schodel11, Peissker20}. 
More details on this extended gas emission are reported in Appendix~\ref{App:gas}. 

Our spectral extraction procedure is designed to reduce the effect of bright stars and extended gas emission at the Galactic center.   
First, to minimize the contribution from bright stars, we create and apply an adaptive mask that tracks the brightest stars and that transmits $\sim$60\% of the aperture area (see Appendix~\ref{App:mask}). 
The extracted spectra are reported in Appendix~\ref{App:ext}.
Second, we apply a continuum subtraction routine to each data cube (see Appendix~\ref{App:cont}) prior to combining the data from all epochs.
The individual epochal spectra are rescaled by dividing by their corresponding unmasked area within the aperture.
The resulting spectra have a typical noise per spectral resolution element of 0.050~mJy (see Table~\ref{tab:obs}).
We then combine the individual spectra using a weighted average where the weight corresponds to the inverse square of the noise. 
The noise of the combined spectrum (0.015~mJy) is lower by a factor of $\sim$2.3 than in the best individual epoch.


\section{Results} \label{sec:res}

\begin{figure*}[htb]
    \centering
    \includegraphics[width=10cm]{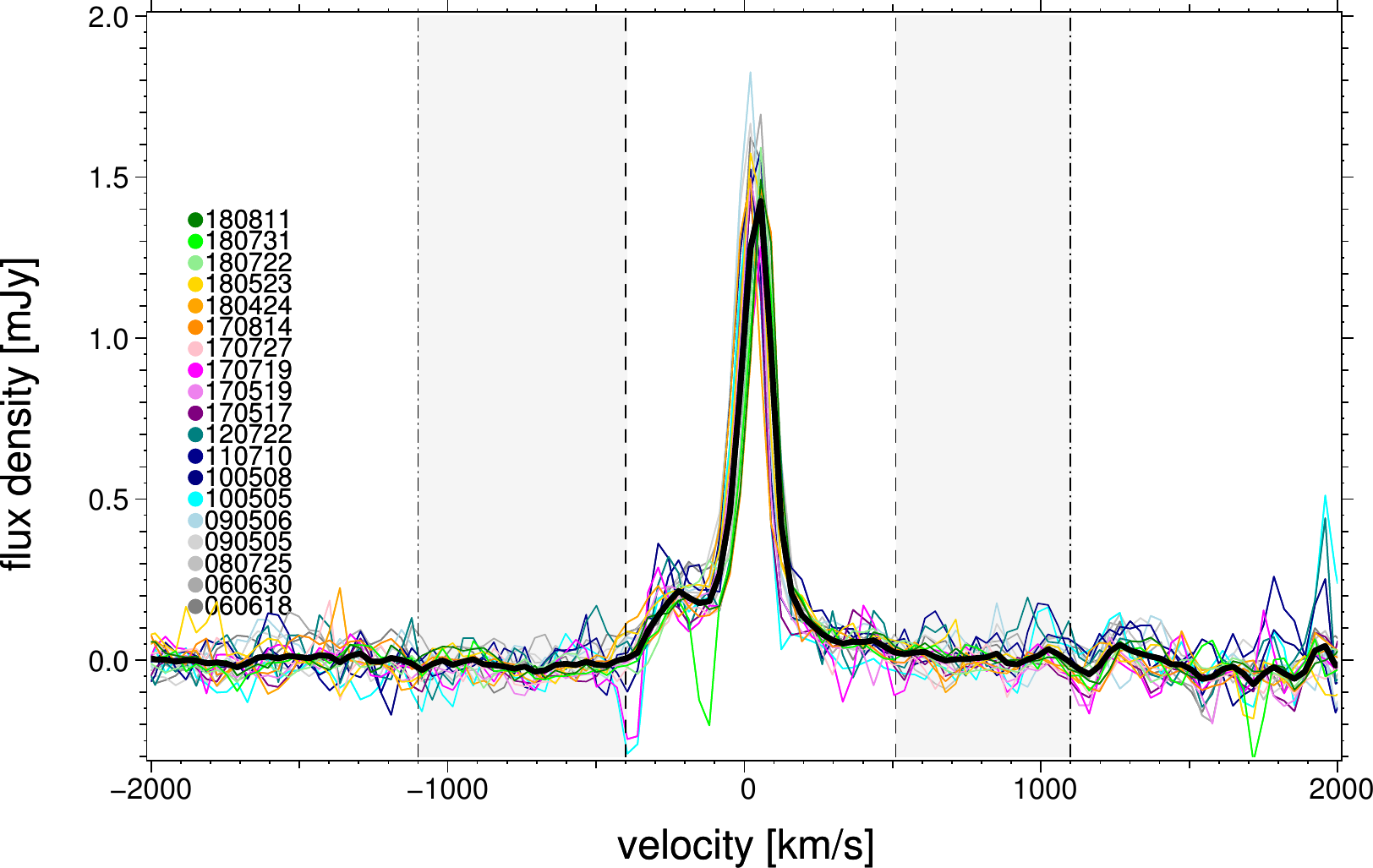}
    \includegraphics[width=10cm]{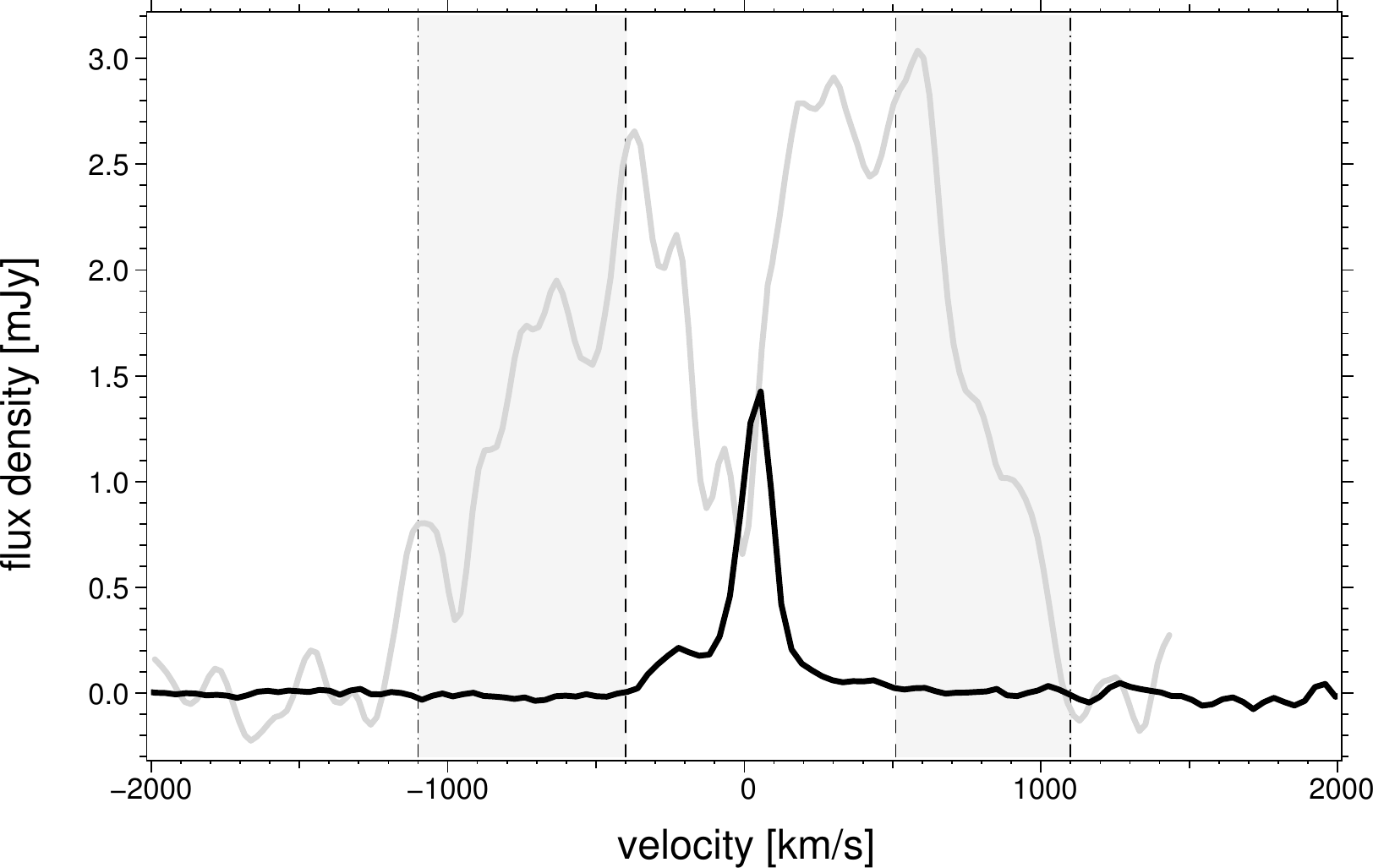}
    \caption{
    $Top:$     Continuum-subtracted spectra extracted over a 0.23'' radius aperture centered on Sgr~A*. 
    The colored lines show the single-epoch spectra normalized over the fraction of aperture used (dates reported in YYMMDD format). 
    The thick black line represents the combined spectrum. There is no broad (several 1000~km/s) line detected.
    $Bottom:$   Observed Br$\gamma$ spectrum (black), compared to the observed H30$\alpha$ spectrum from \cite{Murchikova19} (gray). 
    The flux scale applies to both the radio and NIR emissions.
    At our level of sensitivity, there is no Br$\gamma$ counterpart to the broad, double-peaked H30$\alpha$ emission. 
    The dash-dotted vertical lines represent the velocity range over which the H30$\alpha$ line has been reported.
    The dashed vertical lines shows the much narrower range over which superimposed gas emission is present.
    The shaded area represents our notch filter.
    }
    \label{fig:2lines}
\end{figure*}

Figure~\ref{fig:2lines} (top) shows our resulting combined spectrum along with the individual spectra. 
Subsequent analysis only uses the combined spectrum. 
Figure~\ref{fig:2lines} (bottom) shows the comparison between the observed Br$\gamma$ spectrum (not corrected for extinction) and the H30$\alpha$ spectrum reported by \cite{Murchikova19}.
The only Br$\gamma$ emission that rises above the noise is associated with the superimposed lower-velocity gas that is also observed at larger radii (see Figure~\ref{fig:cont} and Appendix~\ref{App:gas}).
There is no qualitative evidence for the presence of a broad Br$\gamma$ emission line. 
We want to underline that there is no sign of any broad ($>$800~km/s) line in the whole wavelength range covered by the OSIRIS observations ($\sim$12000~km/s). Consequently, this means that there is no counterpart to the even broader radio recombination line recently reported by \cite{Yusef-Zadeh20} ($\sim$4000~km/s).

The most straightforward way to extract an upper limit on the flux of the Br$\gamma$ line, without making any assumptions about the shape or extent of the line profile, is to set the limit at 5 times the root-mean-square noise per resolution element.
The noise per resolution element is calculated over ranges devoid of spectral features (see Appendix~\ref{App:cont}).  
In this way, we find an upper limit on the flux density of  0.074~mJy in a 0.23'' radius aperture 
When corrected for the extinction reported by \cite{Schodel10}, this limit corresponds to 0.71~mJy.
This limit can be used with any hypothetical line profile to derive the total line flux expected for the accretion flow.

To quantify the non-detection of any broad-band emission ($\sim$2000~km/s), we create a spectral filter that is sensitive to the wings of the reported H30$\alpha$ line and cuts out the low velocities at which the superimposed gas is emitting.
Our notch filter extends from a Br$\gamma$ velocity of -1100 to +1100~km/s excluding the low velocities between -400 and +510~km/s (see Figure~\ref{fig:2lines}); this filter contains 15 resolution elements. 

\begin{figure*}[htb]
    \centering
    \includegraphics[width=13cm]{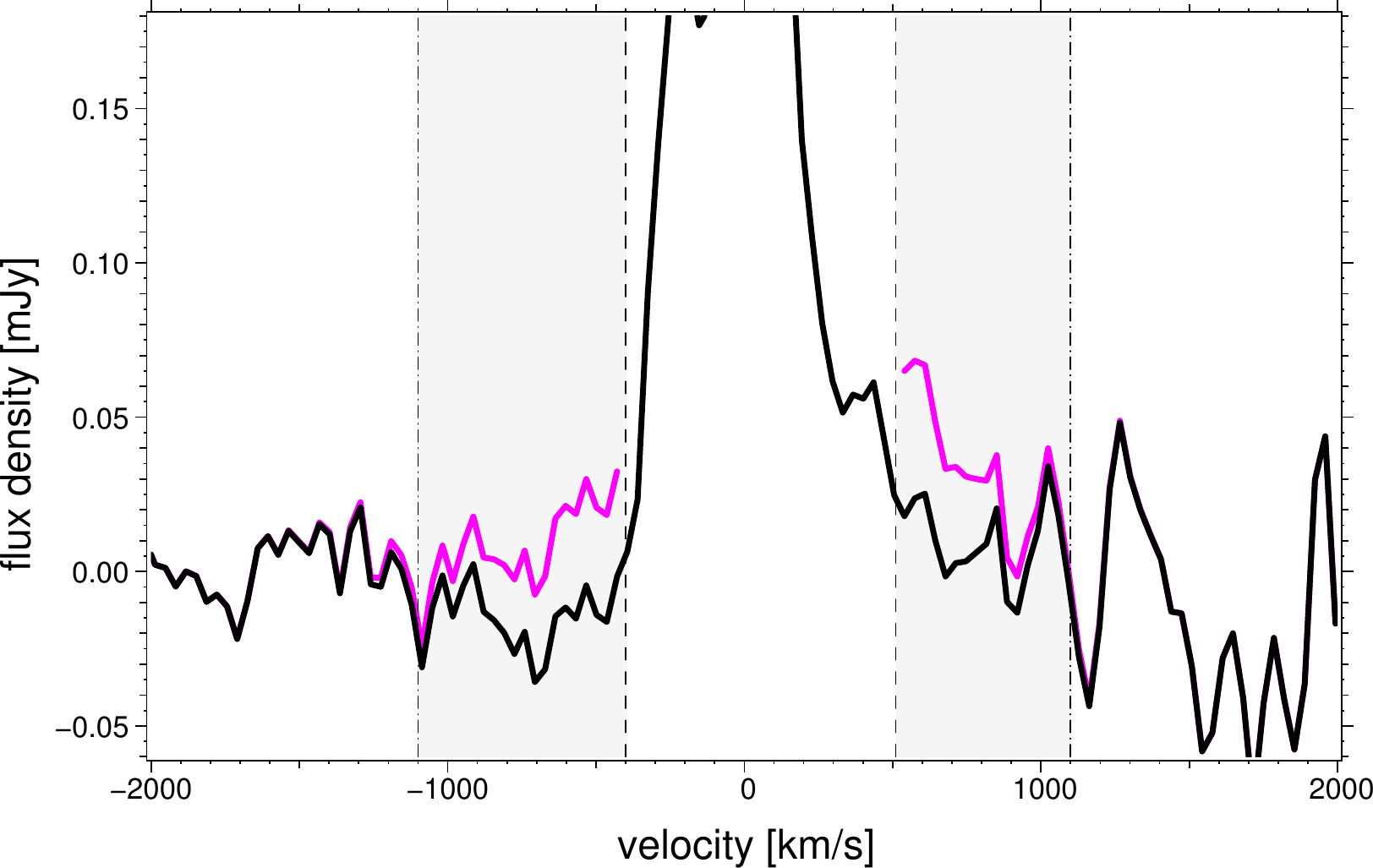}
    \caption{
    Vertically expanded version of the NIR spectrum of Figure~\ref{fig:2lines} (black) together with the spectrum that would correspond to a 5-$\sigma$ detection of a line with the same shape as H30$\alpha$ (only shown in our notch filter, magenta). 
    The shaded area represents the notch filter.
    }
   \label{fig:limits}
\end{figure*}
Using the H30$\alpha$ line profile reported by \cite{Murchikova19}, we proceed next to derive a limit on the Br$\gamma$ line flux by determining our sensitivity to the detection of a line having that profile in our spectrum. 
We do this by modelling the published H30$\alpha$ line profile as a double-peaked Gaussian, and then planting the result of that fit on top of our NIR spectrum.
We only consider the portion of the resulting spectrum in our notch filter. 
A limit to the Br$\gamma$ flux is then determined by identifying the total added flux for which the signal-to-noise ratio of the planted spectrum within the notched filter is greater than 5.
The planted-Gaussian spectrum corresponding to our 5-$\sigma$ limit is shown in Figure~\ref{fig:limits}.
More details can be found in Appendix~\ref{App:plant}. 

Since we are assuming that the Br$\gamma$ profile has the same shape as that of the H30$\alpha$ line, we determine that the notch filter would contain 46\% of the flux.
Hence we find that the upper limit to the total Br$\gamma$ flux (calculated over the whole H30$\alpha$ range, -1100 to +1100~km/s) corresponding to a signal-to-noise ratio of 5 is 2.461$\pm$0.003~10$^{-19}$~W/m$^2$.  
The de-reddened value corresponds to 23.7$\pm$0.03~10$^{-19}$~W/m$^2$. 

We note that, if we don't mask the brightest stars in our spectral extraction, the resulting 5$\sigma$ flux limit is $\sim$15\% higher because of higher photon noise.


\section{Discussion} 
\label{sec:discuss}

In order to compare our flux limit to models of the accretion flow and to the reported H30$\alpha$ line, it is necessary to calculate the corresponding emissivities.  
The total flux integrated over the H30$\alpha$ line (from -1100~km/s to +1100~km/s) is estimated to be 0.3~10$^{-19}$~W/m$^2$ (corrected for extinction).
Our flux limit over the same velocity range\footnote{Note the large factor in the ratios of fluxes versus flux densities for a given velocity range, because of the large difference in the number of Hz per km/s between Br$\gamma$ and H30$\alpha$.} for Br$\gamma$ is 23.7$\pm$0.1~10$^{-19}$~W/m$^2$ (corrected for extinction). 
The corresponding emissivity $\epsilon$ can be calculated as:
\begin{equation}
    \epsilon = \frac{F_\lambda \cdot 4\pi d^2}{V},
\end{equation}
where $F_\lambda$ is the total flux of a given emission line, $d$ is the distance to the Galactic Center (7.97~kpc, \citealt{Do19GR}) and $V$ is the emitting volume.
We consider two geometries for the emitting region: 1) a disk of radius 0.23'' (same as the disk reported by \citealt{Murchikova19}) and height 0.03'' (13\% of the radius), 2) a sphere of radius 0.23''. 
The disk volume occupies roughly a tenth of the sphere volume. 
Using the disk geometries we obtain emissivities for Br$\gamma$ and H30$\alpha$ of $\le$770.9~10$^{-19}$ and 9.8~10$^{-19}$~erg/s/cm$^2$ respectively.
For a sphere geometry we obtain $\le$75.2~10$^{-19}$ and 0.9 ~10$^{-19}$~erg/s/cm$^2$ respectively.

\begin{figure*}
    \centering
    \includegraphics[width=12cm]{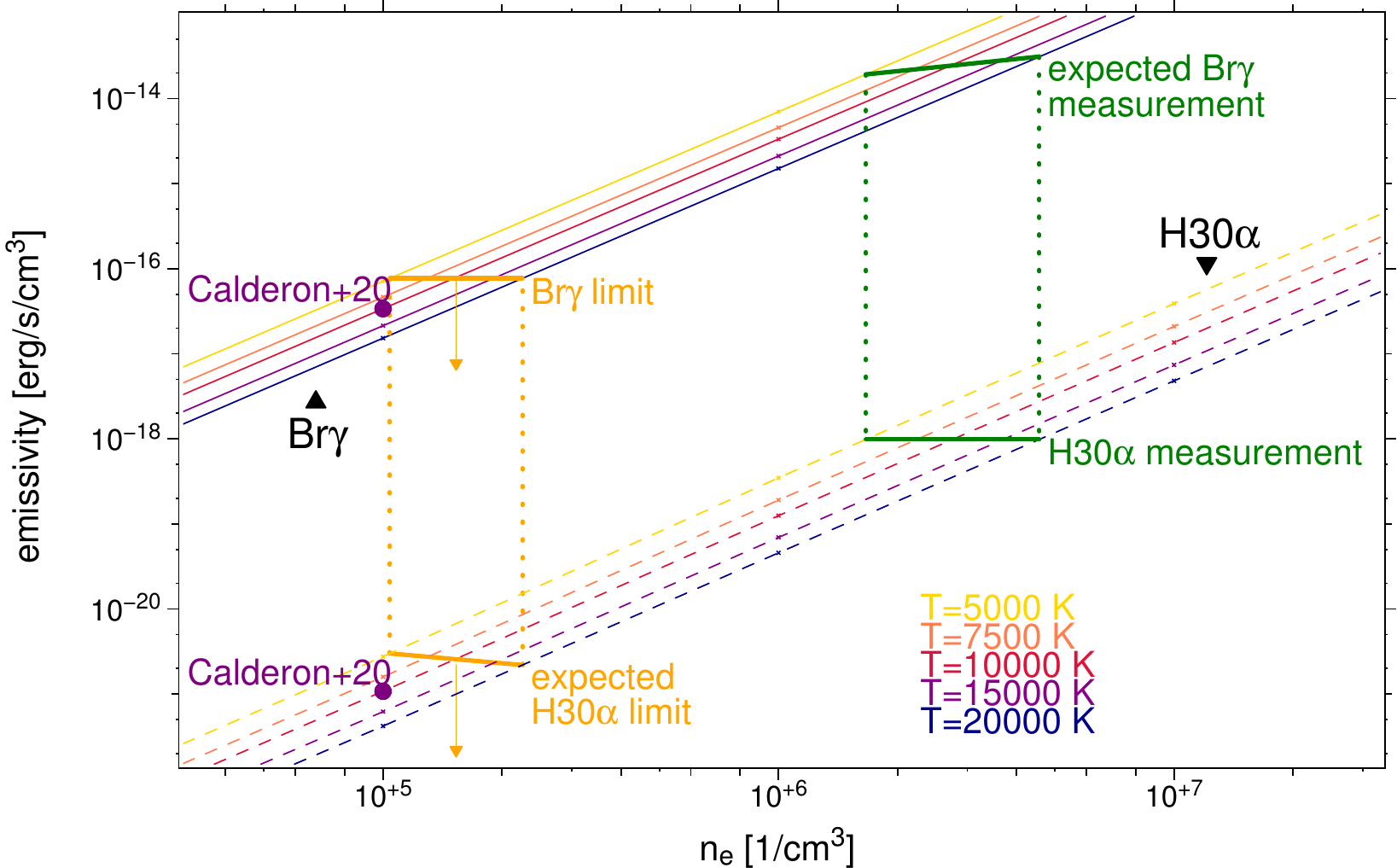}   
    \caption{
    Comparison between the Br$\gamma$ limit, the H30$\alpha$ measurement and the disk characteristics emerging from the model of \citealt{Calderon20} assuming a disk geometry. 
    This plot show emissivity as a function of electron density for a range of temperatures (displayed as groups of lines of different colors) for both Br$\gamma$ (solid lines, upper left group) and H30$\alpha$ (dashed lines, lower right group) from \cite{Hummer87}.
    Small crosses on the lines represent the points where the emissivity was actually calculated by \cite{Hummer87}.
    Br$\gamma$ and H30$\alpha$ emissivities computed from the \cite{Calderon20} disk model density are displayed as purple dots.  
    The H30$\alpha$ measurement by \cite{Murchikova19} leads to an estimated emissivity compatible with a range of densities and is shown as the a green line. 
    The corresponding, predicted Br$\gamma$ emissivity is reported as a green line as well.    
    The same can be done with the Br$\gamma$ flux limits reported as a lines (with arrows to underline that it is purely an upper limit).
    }
    \label{fig:models}
\end{figure*}

To investigate what ranges of temperature and density can give rise to these emissivities, we use the tabulations of \cite{Hummer87}.
Figure~\ref{fig:models} shows the \cite{Hummer87} emissivities as a function of density, where the different colored curves correspond to different temperatures.
We plot two sets of curves: one for each of Br$\gamma$ and  H30$\alpha$.
The emissivities we derive from the total flux of each line are compatible with a range of densities when considering the temperatures used by \cite{Hummer87}. 

Given the the H30$\alpha$ emissivity derived from observations, we can project the compatible densities onto the Br$\gamma$ set of temperature curves, hence determining the expected emissivity in Br$\gamma$ given the observed H30$\alpha$ flux.
The same process can be applied to the Br$\gamma$ upper limit, which implies an upper limit on the expected H30$\alpha$ emission.

We can compare, on the same plot, models of the accretion flow.
For instance, \cite{Calderon20} recently reported simulations of the interactions of stellar winds from stars in the young nuclear cluster that produce a disk on the same scale as the one reported by \cite{Murchikova19}.
According to this model, the temperature and density in the disk are respectively $\sim$10$^{4}$~K and $\sim$10$^{5}$~cm$^{-3}$.

Therefore, we find that: 1) in the case of  an emitting disk, our upper limit on the Br$\gamma$ flux is compatible with the model reported by \cite{Calderon20}, 2) regardless of the modeled emission geometry, the reported H30$\alpha$ emissivity measurement is much higher than that predicted by \cite{Calderon20} and by our Br$\gamma$ limit. 
The width and flux reported by \cite{Yusef-Zadeh20} would lead to similar conclusions.

Our measured Br$\gamma$ flux limits, together with the expected values from reported H30$\alpha$ line and the \citealt{Calderon20} model, are summarized in Table~\ref{tab:limits}.
The Br$\gamma$ flux expected from the H30$\alpha$ observations is several orders of magnitude greater than our reported limit.
This discrepancy between the reported H30$\alpha$ and the upper limit on Br$\gamma$ led \cite{Murchikova19} to propose a maser effect to amplify the H30$\alpha$ emission. 
However, such a population inversion is not expected from the calculations of \cite{Hummer87}, as reported by \cite{Scoville13b} and shown in Figure~\ref{fig:models}.
Although \cite{Hummer87} probe only a limited range of temperatures and densities that might not represent the physical conditions appropriate to the Sgr~A* accretion flow,  we note that, if the gas stays very hot, we should not expect to observe any recombination line, since emissivity declines with temperature (the recombination rate is low at high temperatures because protons are moving too fast).

\begin{table*}
    \centering
    \begin{tabular}{|c|c|}
    \hline
    Measured Br$\gamma$ flux limit based on 5$\times$RMS noise [10$^{-19}$ W/m$^2$] & $\le$72.6 \\
    Measured Br$\gamma$ flux limit based on H30$\alpha$ profile [10$^{-19}$ W/m$^2$] & $\le$ 23.71$\pm$0.03 \\
    \hline
    Expected Br$\gamma$ flux from H30$\alpha$ [10$^{-19}$ W/m$^2$] & 5900 \\
    Expected Br$\gamma$ flux from Calderon et al. model [10$^{-19}$ W/m$^2$] & 10 \\ 
    \hline
    \end{tabular}
    \caption{
    Summary table of Br$\gamma$ flux limits measured in this work and expected fluxes based on the reported H30$\alpha$ line \citep{Murchikova19} and \citealt{Calderon20} model.
    Our measured flux limit from noise corresponds to 5 times the root-mean-square of the noise (multiplied by the expected width of the line, 2200~km/s) whereas the second limit is the one obtained by planting the H30$\alpha$ profile in our spectrum.
    Both expected fluxes are calculated using a disk geometry as previously described. 
    All values are corrected for extinction.
    }
    \label{tab:limits}
\end{table*}

If the level populations of the H30$\alpha$ line are inverted, maser amplification can occur.   \cite{Murchikova19} have invoked this mechanism in order to reproduce the flux densities that they report, given our upper limit to the Br-$\gamma$ line. 
However, the nearby presence of the very bright (several Jansky) point source, Sgr A*, at the center of the putative disk is problematical for the maser hypothesis because the preferred gain path for the maser would always be radial, directed away from Sgr A*.  
In the competition for stimulating new photon emission from the upper state of the H30$\alpha$ transition, the strong emission from Sgr A* would be strongly favored over the relatively weak (few milli-Jansky), isotropically emitted, spontaneous emission in the H30$\alpha$ line from throughout the accretion disk.  
Consequently, if that transition were indeed inverted, the resulting maser-enhanced line radiation would always be directed along a radial ray from Sgr A*.   
Then, rather than observing emission extended along the projected disk and showing a strong velocity difference on opposite sides of the projected disk, as was reported by \cite{Murchikova19}, we should expect to see an unresolved point source of line emission having a velocity width reflecting the line-of-sight velocity gradient in the accretion flow.  
Furthermore, the reported 2000 km s$^{-1}$ width of the line is far larger than could be expected for the radial component of gas moving in quasi-circular orbits in an accretion disk.  
We conclude from these arguments that the maser interpretation of the reported H30$\alpha$ line emission is unlikely.

\section{Conclusions} \label{sec:concl}

Analyzing 13 years of data gathered with the integral field spectrograph OSIRIS at the W.M. Keck Observatory, we extracted the near-infrared spectrum of the immediate vicinity of Sgr~A*. 
Within a radius of 0.23'' we place a 5-sigma upper limit Brackett-$\gamma$ emission line (see Table~\ref{tab:limits} for a summary).
Predictions drawn from future modeling of the accretion flow will need to be constrained by this result.

Our limit on Br$\gamma$ emission is unexpected given the reported millimeter-wavelength detection of the higher-level transition H30$\alpha$ \citep{Murchikova19} within the same region. 
Given our limit on Br$\gamma$, the reported excitation of H30$\alpha$ would need to deviate dramatically from current theoretical expectations based on the departure coefficients computed by \citet{Hummer87}.
However, it remains possible that the range of physical conditions applicable to the Sgr~A* accretion flow is outside of those they considered.
Nevertheless, if strong maser amplification can in principle allow for much stronger H30$\alpha$ emission, we would expect to observe it as an unresolved point source rather than a disk.

\acknowledgments
Support for this work was provided by NSF AAG grant AST-1412615, Jim and Lori Keir, the W. M. Keck Observatory Keck Visiting Scholar program, the Gordon and Betty Moore Foundation, the Heising-Simons Foundation, and Howard and Astrid Preston. A. M. G. acknowledges support from her Lauren B. Leichtman and Arthur E. Levine Endowed Astronomy Chair. 
The W. M. Keck Observatory is operated as a scientific partnership among the California Institute of Technology, the University of California, and the National Aeronautics and Space Administration. The Observatory was made possible by the generous financial support of the W. M. Keck Foundation. 
The authors wish to recognize and acknowledge the very significant cultural role and reverence that the summit of Maunakea has always had within the indigenous Hawaiian community. We are most fortunate to have the opportunity to conduct observations from this mountain.

\bibliography{broad_line}

\begin{thebibliography}{}
\expandafter\ifx\csname natexlab\endcsname\relax\def\natexlab#1{#1}\fi
\providecommand{\url}[1]{\href{#1}{#1}}

\bibitem[{{Baganoff} {et~al.}(2003){Baganoff}, {Maeda}, {Morris}, {Bautz},
  {Brandt}, {Cui}, {Doty}, {Feigelson}, {Garmire}, {Pravdo}, {Ricker}, \&
  {Townsley}}]{Baganoff03}
{Baganoff}, F.~K., {Maeda}, Y., {Morris}, M., {et~al.} 2003, \apj, 591, 891

\bibitem[{{Boehle} {et~al.}(2016){Boehle}, {Ghez}, {Sch{\"o}del}, {Meyer},
  {Yelda}, {Albers}, {Martinez}, {Becklin}, {Do}, {Lu}, {Matthews}, {Morris},
  {Sitarski}, \& {Witzel}}]{Boehle16}
{Boehle}, A., {Ghez}, A.~M., {Sch{\"o}del}, R., {et~al.} 2016, \apj, 830, 17

\bibitem[{{Calder{\'o}n} {et~al.}(2020){Calder{\'o}n}, {Cuadra}, {Schartmann},
  {Burkert}, \& {Russell}}]{Calderon20}
{Calder{\'o}n}, D., {Cuadra}, J., {Schartmann}, M., {Burkert}, A., \&
  {Russell}, C. M.~P. 2020, \apjl, 888, L2

\bibitem[{{Ciurlo} {et~al.}(2020){Ciurlo}, {Campbell}, {Morris}, {Do}, {Ghez},
  {Hees}, {Sitarski}, {Kosmo O'Neil}, {Chu}, {Martinez}, {Naoz}, \&
  {Stephan}}]{Ciurlo20}
{Ciurlo}, A., {Campbell}, R.~D., {Morris}, M.~R., {et~al.} 2020, \nat, 577, 337

\bibitem[{{Clavel} {et~al.}(2013){Clavel}, {Terrier}, {Goldwurm}, {Morris},
  {Ponti}, {Soldi}, \& {Trap}}]{Clavel13}
{Clavel}, M., {Terrier}, R., {Goldwurm}, A., {et~al.} 2013, \aap, 558, A32

\bibitem[{{Davies}(2007)}]{Davies07}
{Davies}, R.~I. 2007, \mnras, 375, 1099

\bibitem[{{Do} {et~al.}(2009){Do}, {Ghez}, {Morris}, {Lu}, {Matthews}, {Yelda},
  \& {Larkin}}]{Do09}
{Do}, T., {Ghez}, A.~M., {Morris}, M.~R., {et~al.} 2009, \apj, 703, 1323

\bibitem[{{Do} {et~al.}(2019{\natexlab{a}}){Do}, {Witzel}, {Gautam}, {Chen},
  {Ghez}, {Morris}, {Becklin}, {Ciurlo}, {Hosek}, {Martinez}, {Matthews},
  {Sakai}, \& {Sch{\"o}del}}]{Do19}
{Do}, T., {Witzel}, G., {Gautam}, A.~K., {et~al.} 2019{\natexlab{a}}, \apjl,
  882, L27

\bibitem[{{Do} {et~al.}(2019{\natexlab{b}}){Do}, {Hees}, {Ghez}, {Martinez},
  {Chu}, {Jia}, {Sakai}, {Lu}, {Gautam}, {O'Neil}, {Becklin}, {Morris},
  {Matthews}, {Nishiyama}, {Campbell}, {Chappell}, {Chen}, {Ciurlo},
  {Dehghanfar}, {Gallego-Cano}, {Kerzendorf}, {Lyke}, {Naoz}, {Saida},
  {Sch{\"o}del}, {Takahashi}, {Takamori}, {Witzel}, \& {Wizinowich}}]{Do19GR}
{Do}, T., {Hees}, A., {Ghez}, A., {et~al.} 2019{\natexlab{b}}, Science, 365,
  664

\bibitem[{{Eckart} \& {Genzel}(1997)}]{Eckart97}
{Eckart}, A., \& {Genzel}, R. 1997, \mnras, 284, 576

\bibitem[{{Ghez} {et~al.}(1998){Ghez}, {Klein}, {Morris}, \&
  {Becklin}}]{Ghez98}
{Ghez}, A.~M., {Klein}, B.~L., {Morris}, M., \& {Becklin}, E.~E. 1998, \apj,
  509, 678

\bibitem[{{Ghez} {et~al.}(2000){Ghez}, {Morris}, {Becklin}, {Tanner}, \&
  {Kremenek}}]{Ghez00}
{Ghez}, A.~M., {Morris}, M., {Becklin}, E.~E., {Tanner}, A., \& {Kremenek}, T.
  2000, \nat, 407, 349

\bibitem[{{Ghez} {et~al.}(2008){Ghez}, {Salim}, {Weinberg}, {Lu}, {Do}, {Dunn},
  {Matthews}, {Morris}, {Yelda}, {Becklin}, {Kremenek}, {Milosavljevic}, \&
  {Naiman}}]{Ghez08}
{Ghez}, A.~M., {Salim}, S., {Weinberg}, N.~N., {et~al.} 2008, \apj, 689, 1044

\bibitem[{{Hummer} \& {Storey}(1987)}]{Hummer87}
{Hummer}, D.~G., \& {Storey}, P.~J. 1987, \mnras, 224, 801

\bibitem[{{Larkin} {et~al.}(2006){Larkin}, {Barczys}, {Krabbe}, {Adkins},
  {Aliado}, {Amico}, {Brims}, {Campbell}, {Canfield}, {Gasaway}, {Honey},
  {Iserlohe}, {Johnson}, {Kress}, {LaFreniere}, {Lyke}, {Magnone}, {Magnone},
  {McElwain}, {Moon}, {Quirrenbach}, {Skulason}, {Song}, {Spencer}, {Weiss}, \&
  {Wright}}]{Larkin06}
{Larkin}, J., {Barczys}, M., {Krabbe}, A., {et~al.} 2006, in \procspie, Vol.
  6269, Society of Photo-Optical Instrumentation Engineers (SPIE) Conference
  Series, 62691A

\bibitem[{{Murchikova} {et~al.}(2019){Murchikova}, {Phinney}, {Pancoast}, \&
  {Blandford}}]{Murchikova19}
{Murchikova}, E.~M., {Phinney}, E.~S., {Pancoast}, A., \& {Blandford}, R.~D.
  2019, \nat, 570, 83

\bibitem[{{Pei{\ss}ker} {et~al.}(2020){Pei{\ss}ker}, {Eckart}, {Sabha},
  {Zaja{\v{c}}ek}, \& {Bhat}}]{Peissker20}
{Pei{\ss}ker}, F., {Eckart}, A., {Sabha}, N.~B., {Zaja{\v{c}}ek}, M., \&
  {Bhat}, H. 2020, \apj, 897, 28

\bibitem[{{Ressler} {et~al.}(2020){Ressler}, {Quataert}, \&
  {Stone}}]{Ressler20}
{Ressler}, S.~M., {Quataert}, E., \& {Stone}, J.~M. 2020, \mnras, 492, 3272

\bibitem[{{Sch{\"o}del} {et~al.}(2011){Sch{\"o}del}, {Morris}, {Muzic},
  {Alberdi}, {Meyer}, {Eckart}, \& {Gezari}}]{Schodel11}
{Sch{\"o}del}, R., {Morris}, M.~R., {Muzic}, K., {et~al.} 2011, \aap, 532, A83

\bibitem[{{Sch{\"o}del} {et~al.}(2010){Sch{\"o}del}, {Najarro}, {Muzic}, \&
  {Eckart}}]{Schodel10}
{Sch{\"o}del}, R., {Najarro}, F., {Muzic}, K., \& {Eckart}, A. 2010, \aap, 511,
  A18

\bibitem[{{Sch{\"o}del} {et~al.}(2002){Sch{\"o}del}, {Ott}, {Genzel},
  {Hofmann}, {Lehnert}, {Eckart}, {Mouawad}, {Alexander}, {Reid}, {Lenzen},
  {Hartung}, {Lacombe}, {Rouan}, {Gendron}, {Rousset}, {Lagrange}, {Brandner},
  {Ageorges}, {Lidman}, {Moorwood}, {Spyromilio}, {Hubin}, \&
  {Menten}}]{Schodel02}
{Sch{\"o}del}, R., {Ott}, T., {Genzel}, R., {et~al.} 2002, \nat, 419, 694

\bibitem[{{Scoville} \& {Murchikova}(2013)}]{Scoville13b}
{Scoville}, N., \& {Murchikova}, L. 2013, \apj, 779, 75

\bibitem[{{Terrier} {et~al.}(2018){Terrier}, {Clavel}, {Soldi}, {Goldwurm},
  {Ponti}, {Morris}, \& {Chuard}}]{Terrier18}
{Terrier}, R., {Clavel}, M., {Soldi}, S., {et~al.} 2018, \aap, 612, A102

\bibitem[{{Witzel} {et~al.}(2012){Witzel}, {Eckart}, {Bremer}, {Zamaninasab},
  {Shahzamanian}, {Valencia-S.}, {Sch{\"o}del}, {Karas}, {Lenzen}, {Marchili},
  {Sabha}, {Garcia-Marin}, {Buchholz}, {Kunneriath}, \&
  {Straubmeier}}]{Witzel12}
{Witzel}, G., {Eckart}, A., {Bremer}, M., {et~al.} 2012, \apjs, 203, 18

\bibitem[{{Wizinowich} {et~al.}(2006){Wizinowich}, {Le Mignant}, {Bouchez},
  {Campbell}, {Chin}, {Contos}, {van Dam}, {Hartman}, {Johansson}, {Lafon},
  {Lewis}, {Stomski}, {Summers}, {Brown}, {Danforth}, {Max}, \&
  {Pennington}}]{Wizinowich06}
{Wizinowich}, P.~L., {Le Mignant}, D., {Bouchez}, A.~H., {et~al.} 2006, \pasp,
  118, 297

\bibitem[{{Yuan} \& {Narayan}(2014)}]{Yaun14}
{Yuan}, F., \& {Narayan}, R. 2014, \araa, 52, 529

\bibitem[{{Yusef-Zadeh} {et~al.}(1990){Yusef-Zadeh}, {Morris}, \&
  {Ekers}}]{Yusef-Zadeh90}
{Yusef-Zadeh}, F., {Morris}, M., \& {Ekers}, R.~D. 1990, \nat, 348, 45

\bibitem[{{Yusef-Zadeh} {et~al.}(2020){Yusef-Zadeh}, {Royster}, {Wardle},
  {Cotton}, {Kunneriath}, {Heywood}, \& {Michail}}]{Yusef-Zadeh20}
{Yusef-Zadeh}, F., {Royster}, M., {Wardle}, M., {et~al.} 2020, \mnras, 499,
  3909

\end{thebibliography}
\bibliographystyle{aasjournal}

\newpage
\appendix

\section{Star mask}
\label{App:mask}
For each epoch we mask a region around the brightest stars by applying a continuum cut. 
Specifically, if the total flux in the wavelength range between 2.1343 and 2.1447~$\mu$m, a region outside our notch filter and devoid of emission features, is higher than a threshold (1.5~mJy) the pixel is masked.
This continuum cut roughly corresponds to masking the brightest stars (K-magnitude$<$16).
The masking serves two purposes: 1) it reduces the noise added by the extended point spread functions of the bright stars and 2) it reduces the confusion potentially caused by Br$\gamma$ absorption in the stellar photospheres of luminous young stars.

\section{Spectral extraction}
\label{App:ext}
\begin{figure}
    \centering
    \includegraphics[width=10cm]{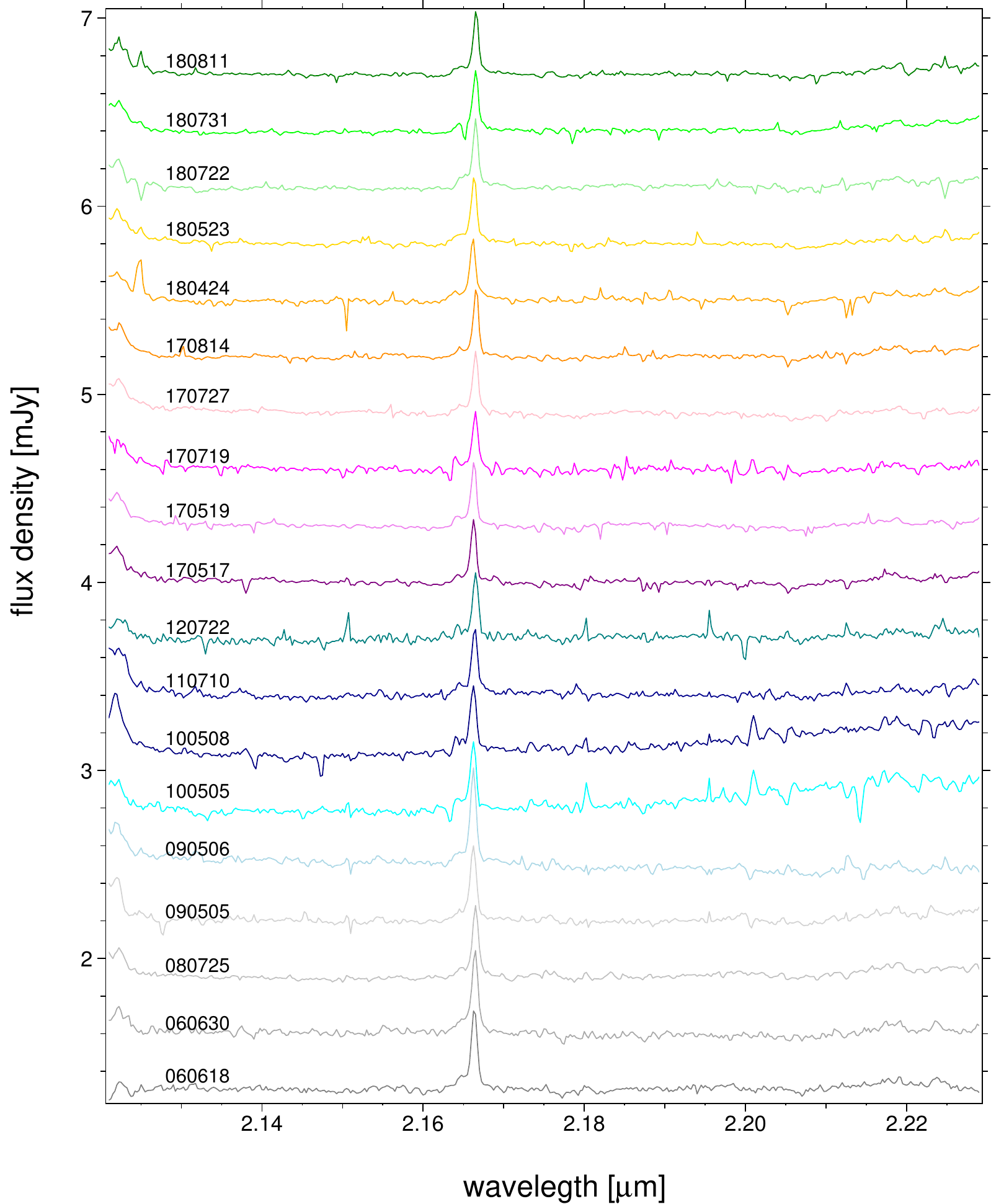}   
    \caption{
    Spectra extracted over a 0.23'' radius aperture centered on Sgr~A* for each epoch (before continuum-subtraction). 
    Each spectrum has been shifted by an arbitrary constant in order to separate and order them from most recent (top) to oldest (bottom) epoch (dates reported in YYMMDD format).
    The spectra with the original continuum levels are displayed in Figure~\ref{fig:wholeall}.
    The central emission line is associated with gas superimposed along the line of sight.
    }
    \label{fig:wholeallr}
\end{figure}
Figure~\ref{fig:wholeallr} shows the spectra extracted over the masked aperture for each epoch.
The spectral characteristics vary from epoch to epoch, showing different slopes. This is due to the fact that we are considering a very small aperture near Sgr~A* through which many stars orbit on short timescales; from year to year different stars enter or exit the aperture.
This happens even though we have masked the brightest stars in each epoch; the many fainter stars in the region still contribute weakly to the extracted spectrum.
The stars in the region are mostly A and B stars, primarily the so-called S-stars orbiting Sgr A* \citep{Eckart97, Ghez98}. 
The late B and A stars have Br$\gamma$ absorption which could potentially influence our measurements, although
most of the S-stars have radial velocities within $\pm$500~km/s, and therefore are not in the range we are considering for our limit on the accretion flow.
Furthermore, for such stars, which have magnitudes above 16, the flux density in the Br$\gamma$ line is expected to be well below the noise of our spectra.

\section{Continuum subtraction and noise estimation}
\label{App:cont}
\begin{figure}[htb]
    \centering
    \includegraphics[width=4.4cm]{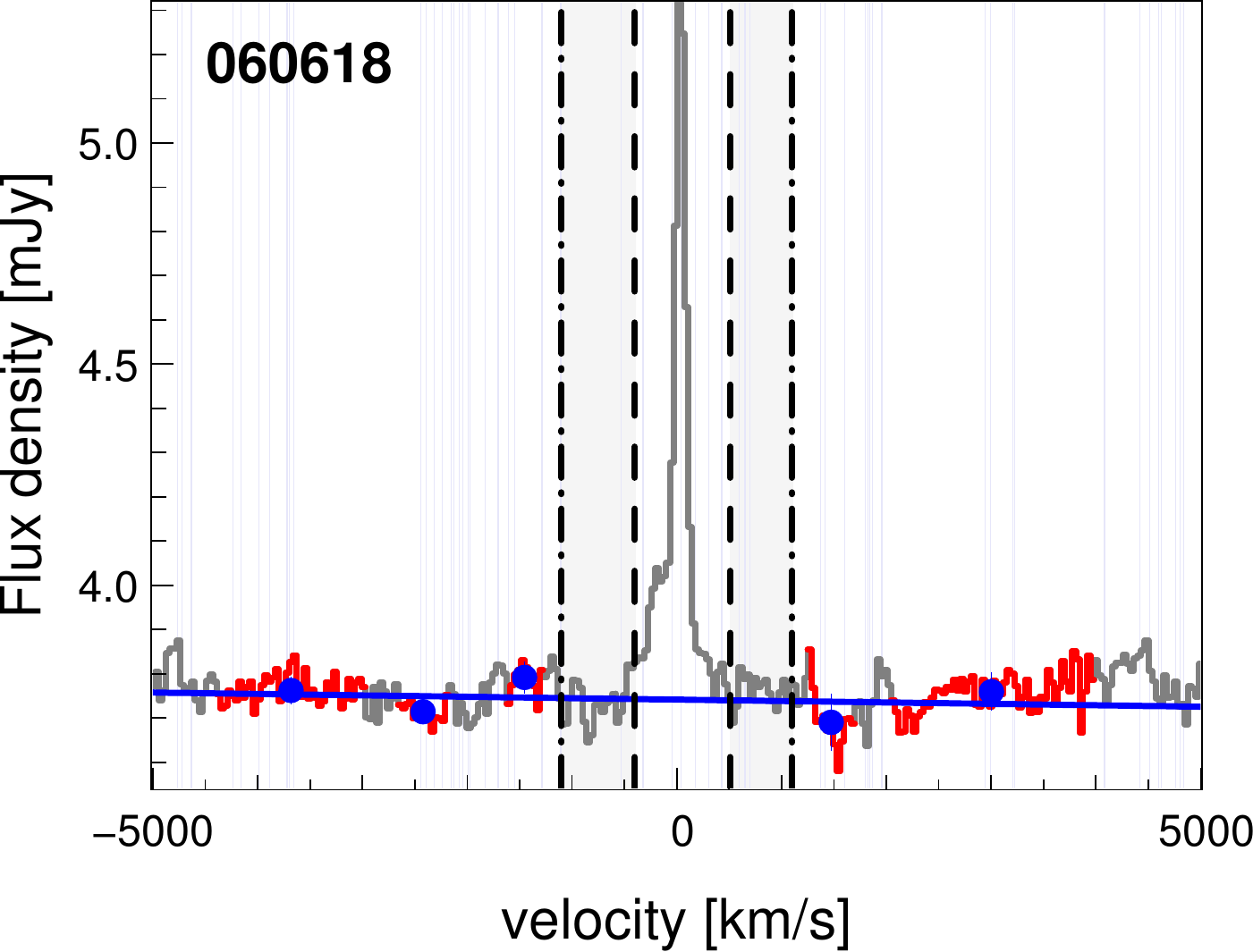}
    \includegraphics[width=4.4cm]{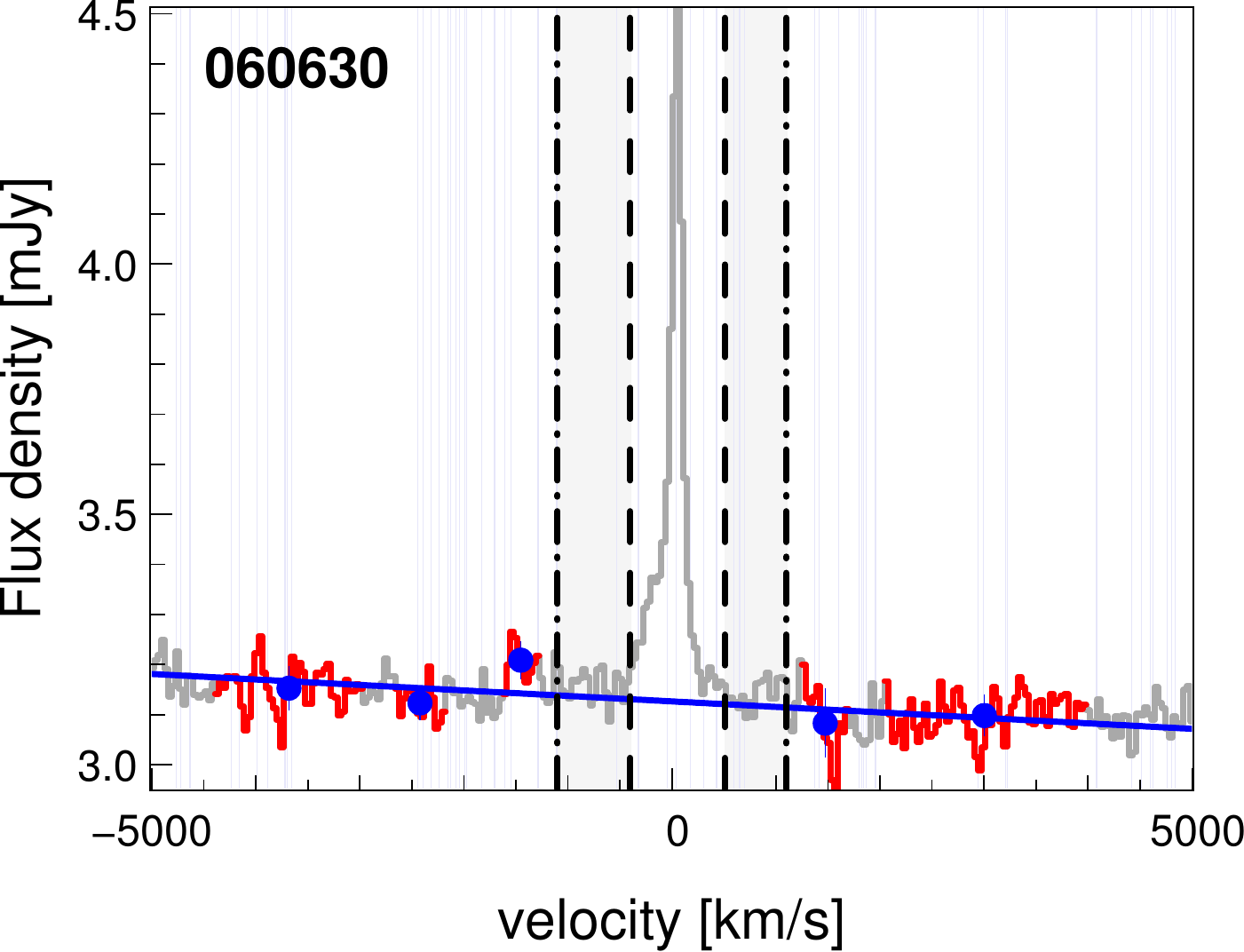}
    \includegraphics[width=4.4cm]{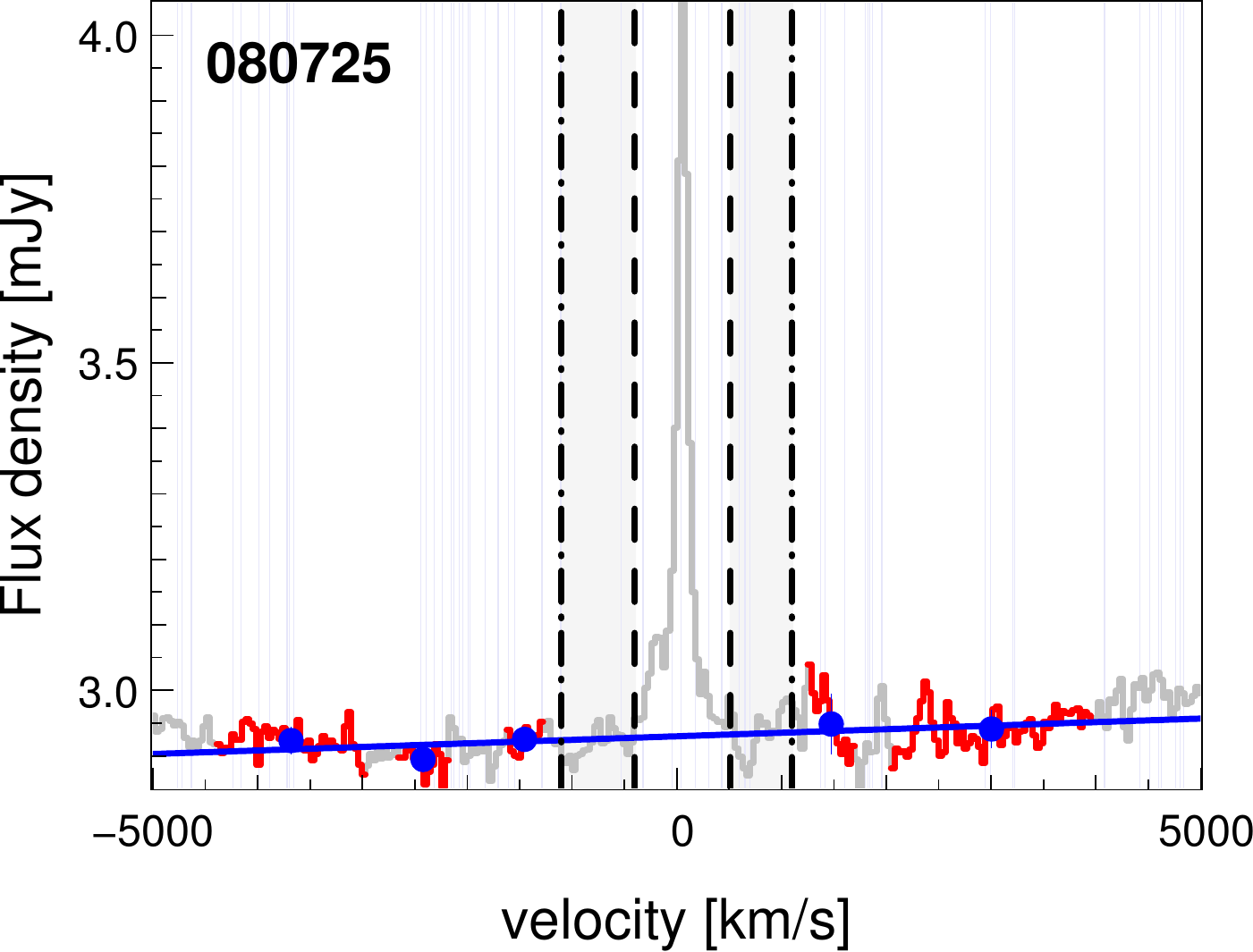}
    \includegraphics[width=4.4cm]{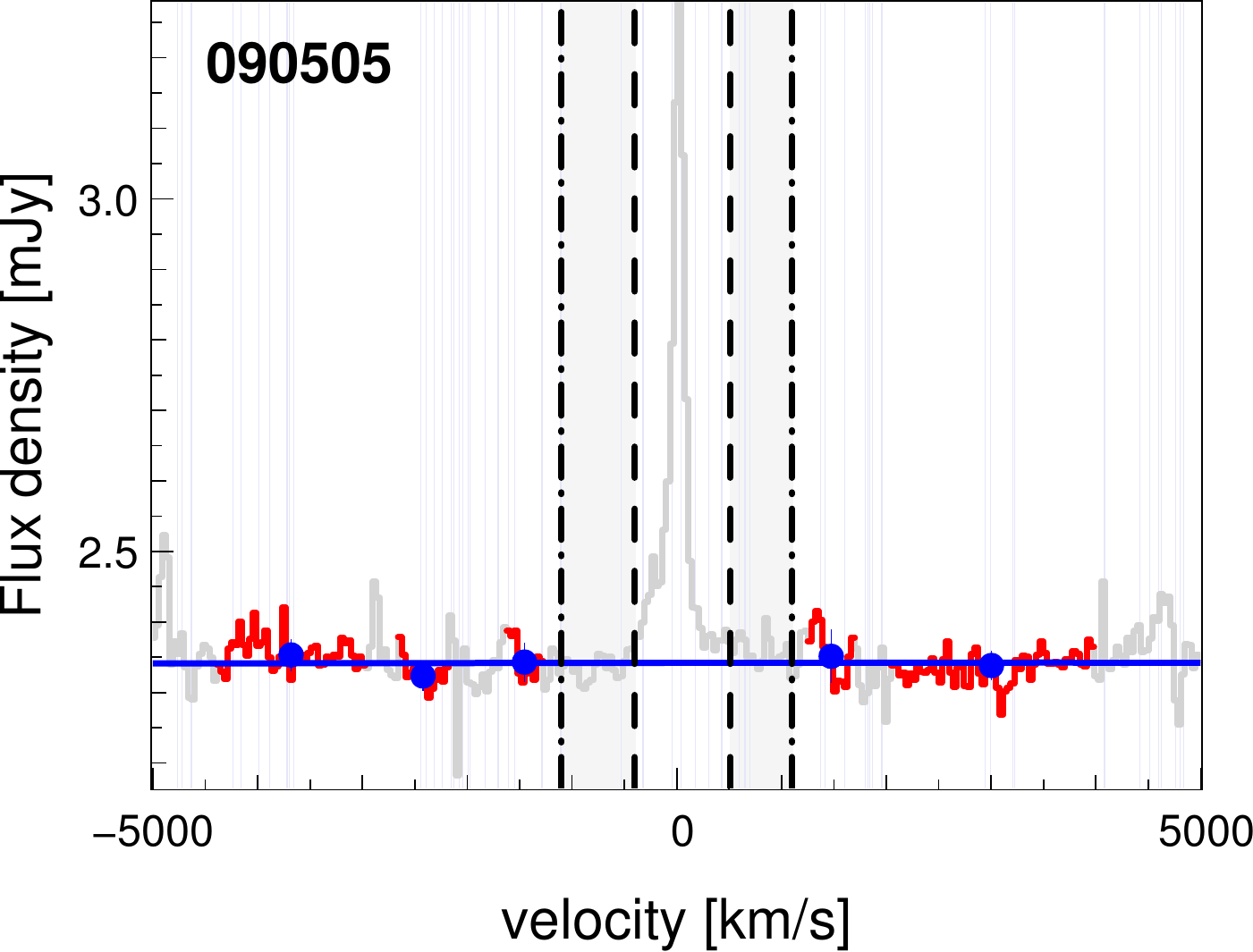}
    \includegraphics[width=4.4cm]{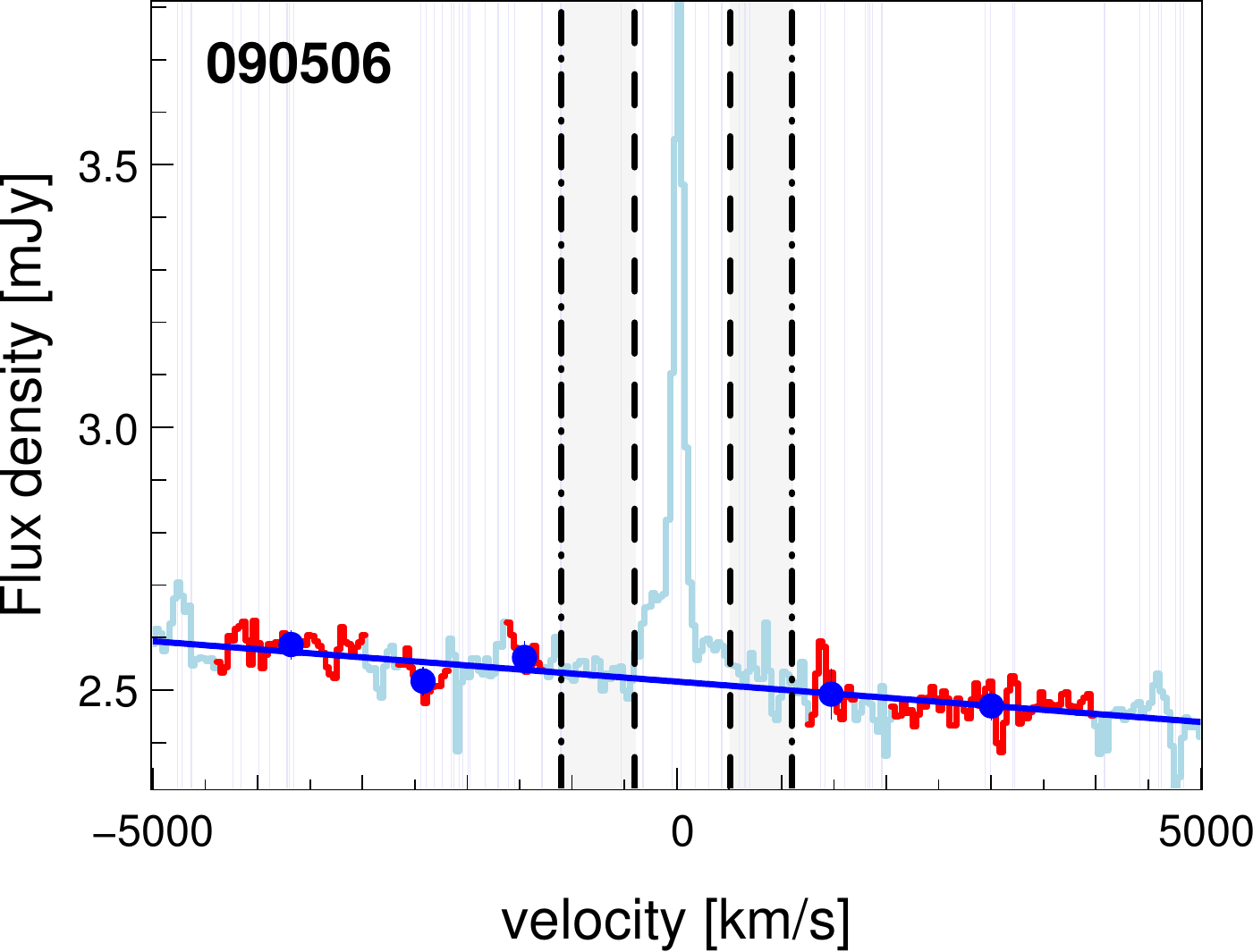}
    \includegraphics[width=4.4cm]{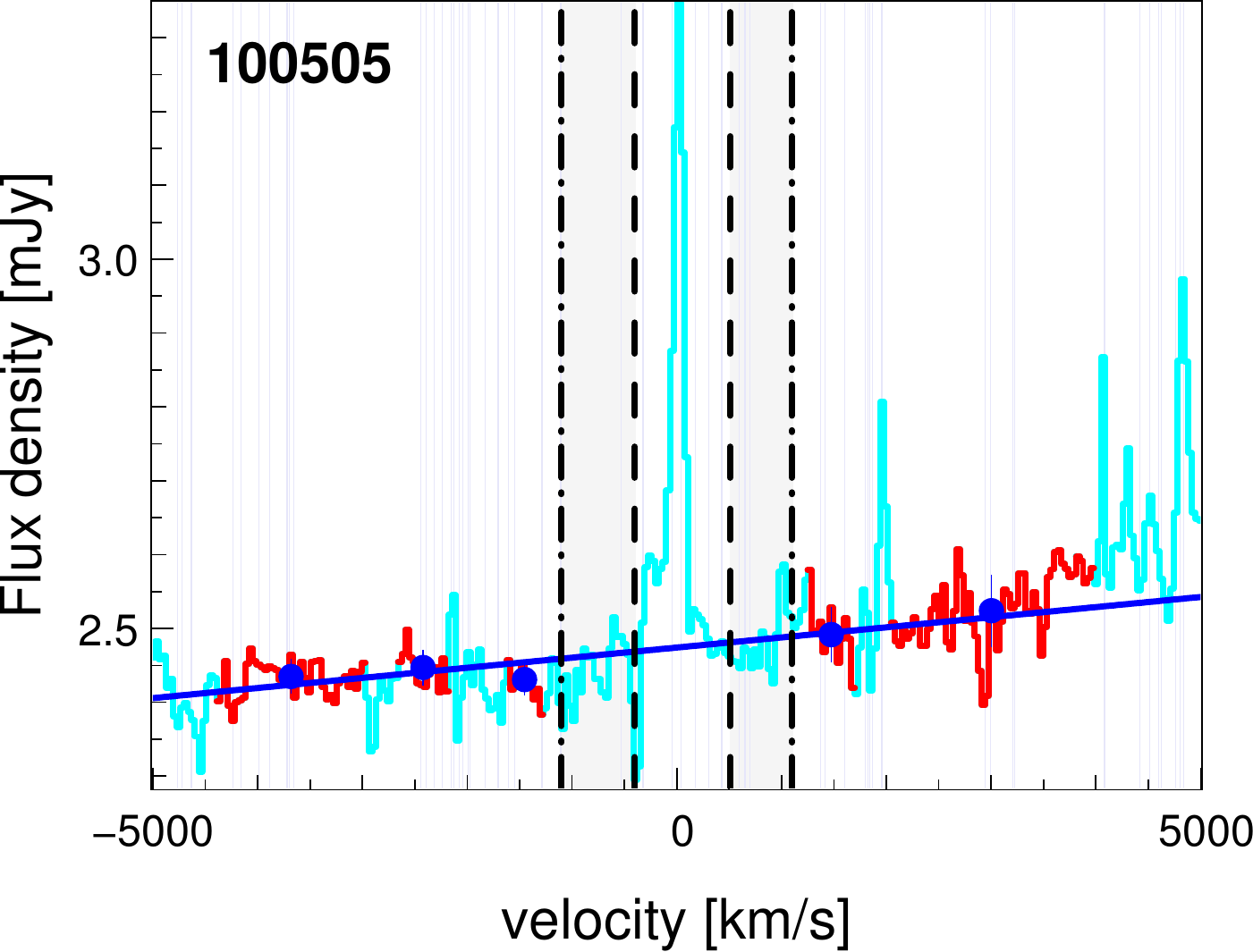}
    \includegraphics[width=4.4cm]{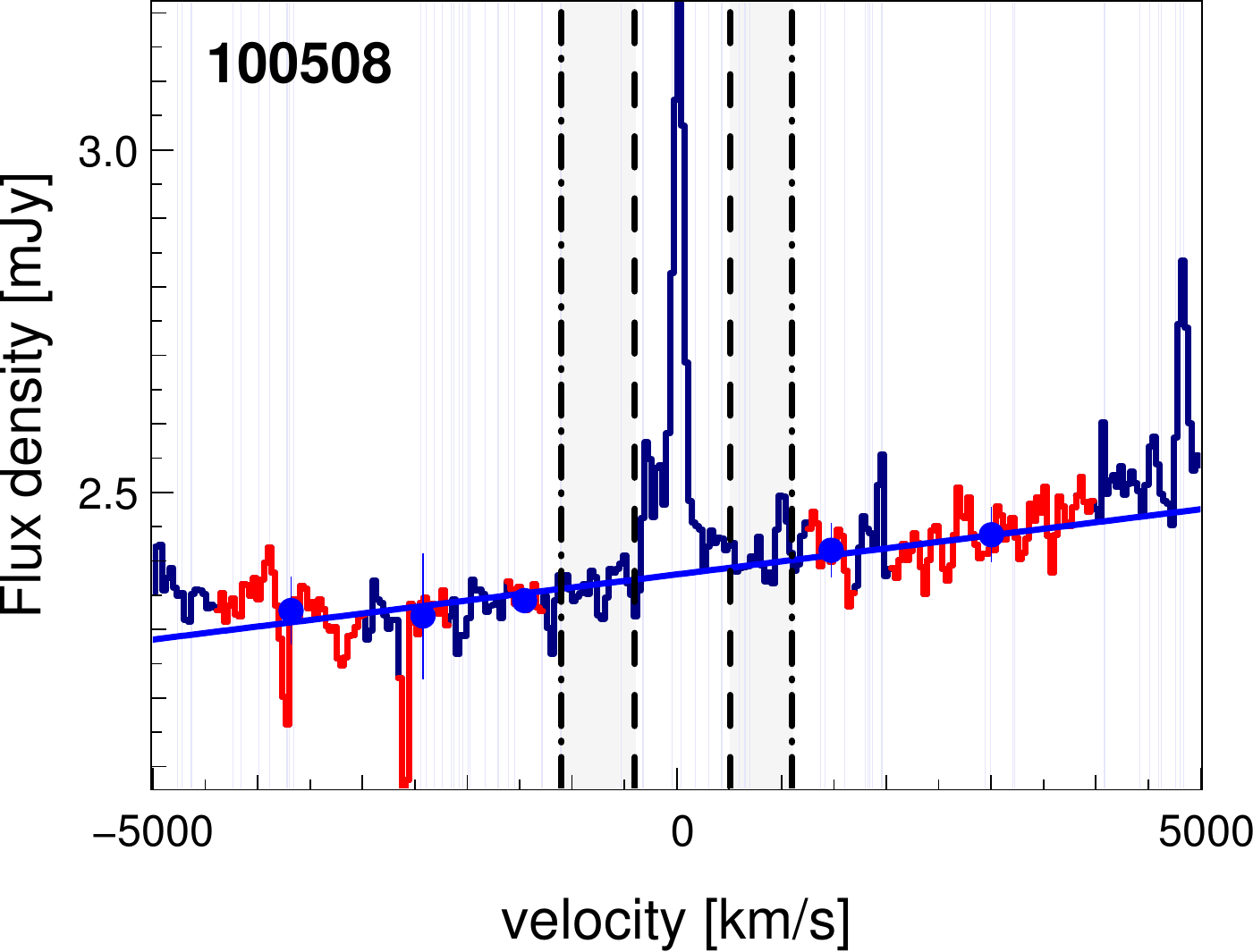}
    \includegraphics[width=4.4cm]{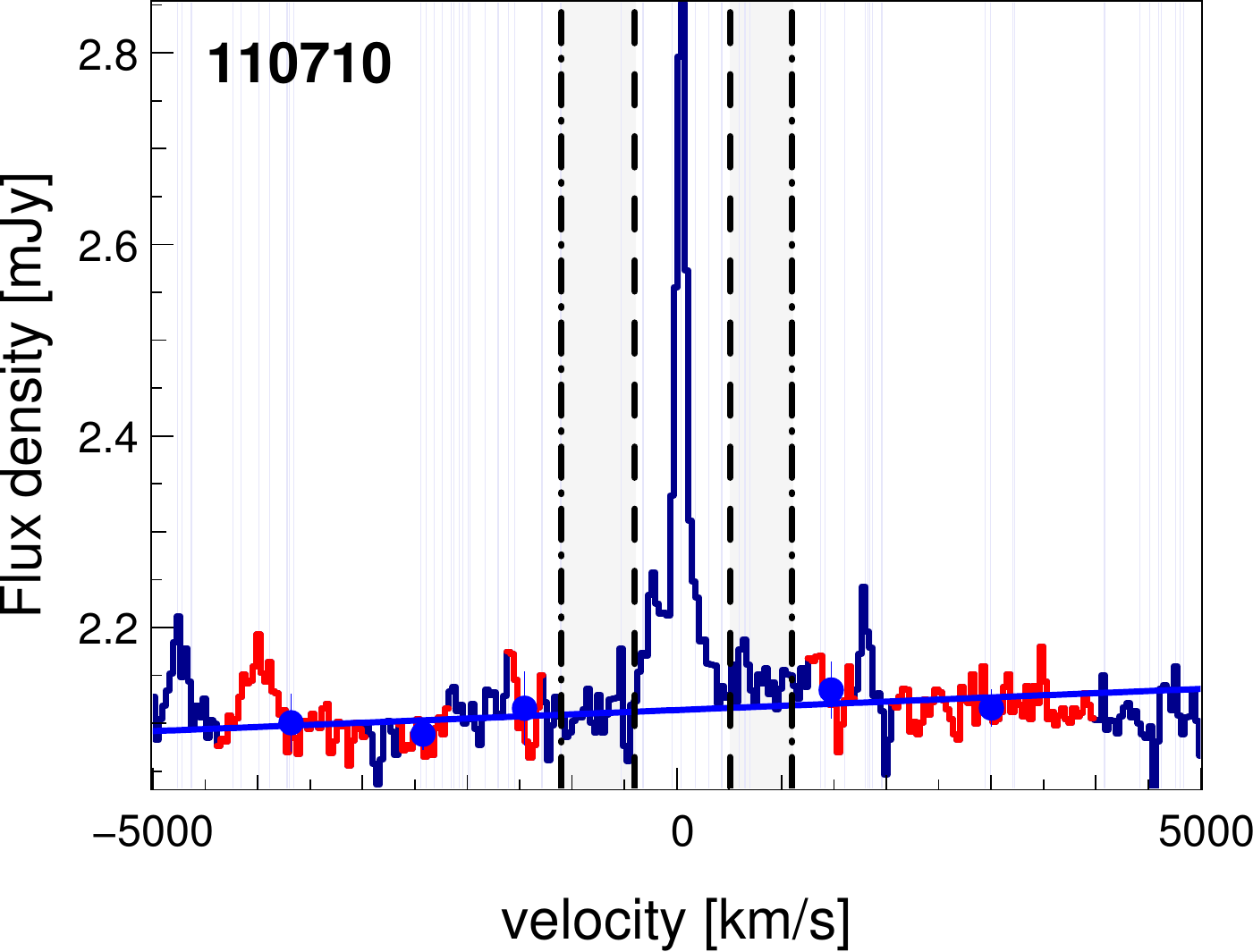}
    \includegraphics[width=4.4cm]{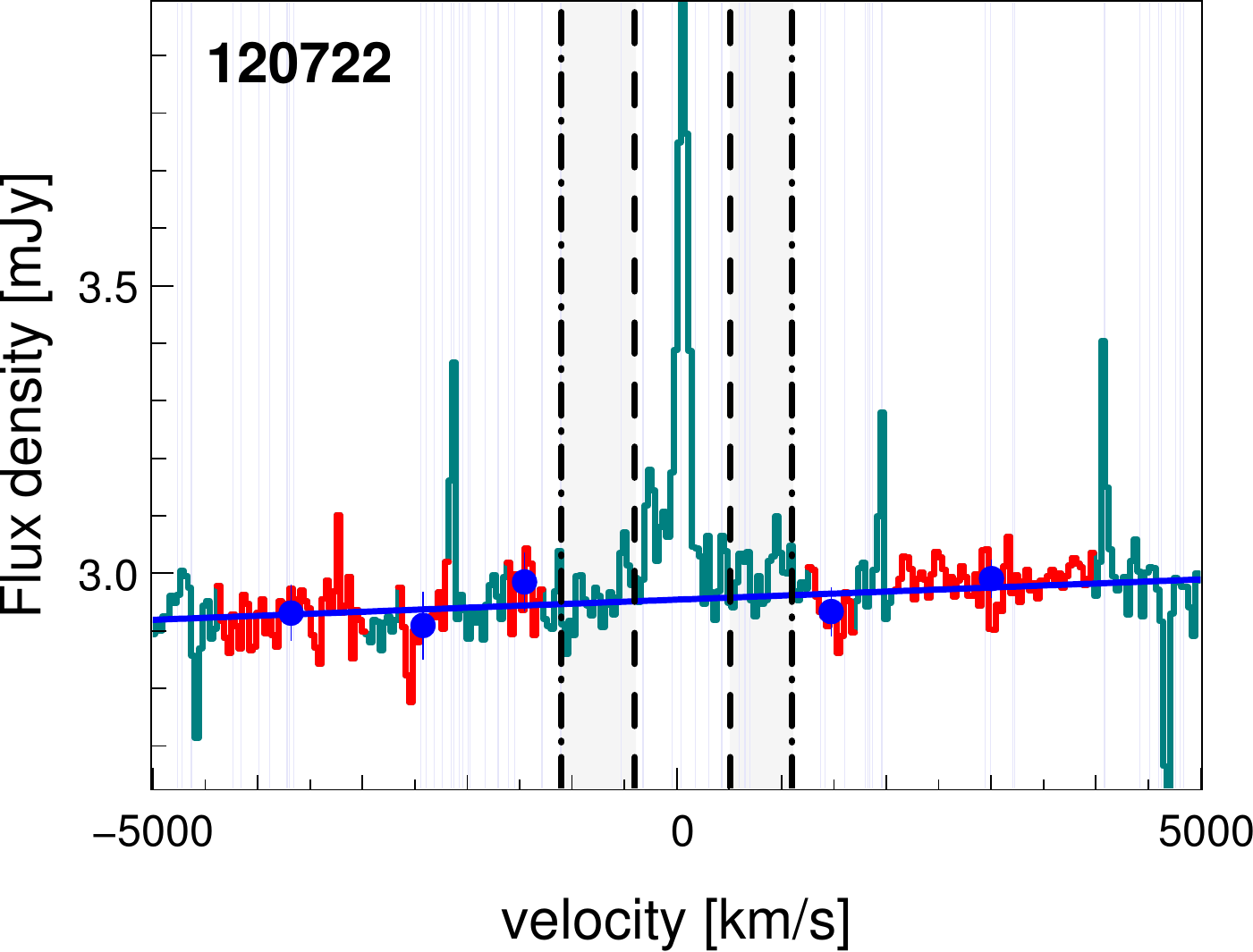}
    \includegraphics[width=4.4cm]{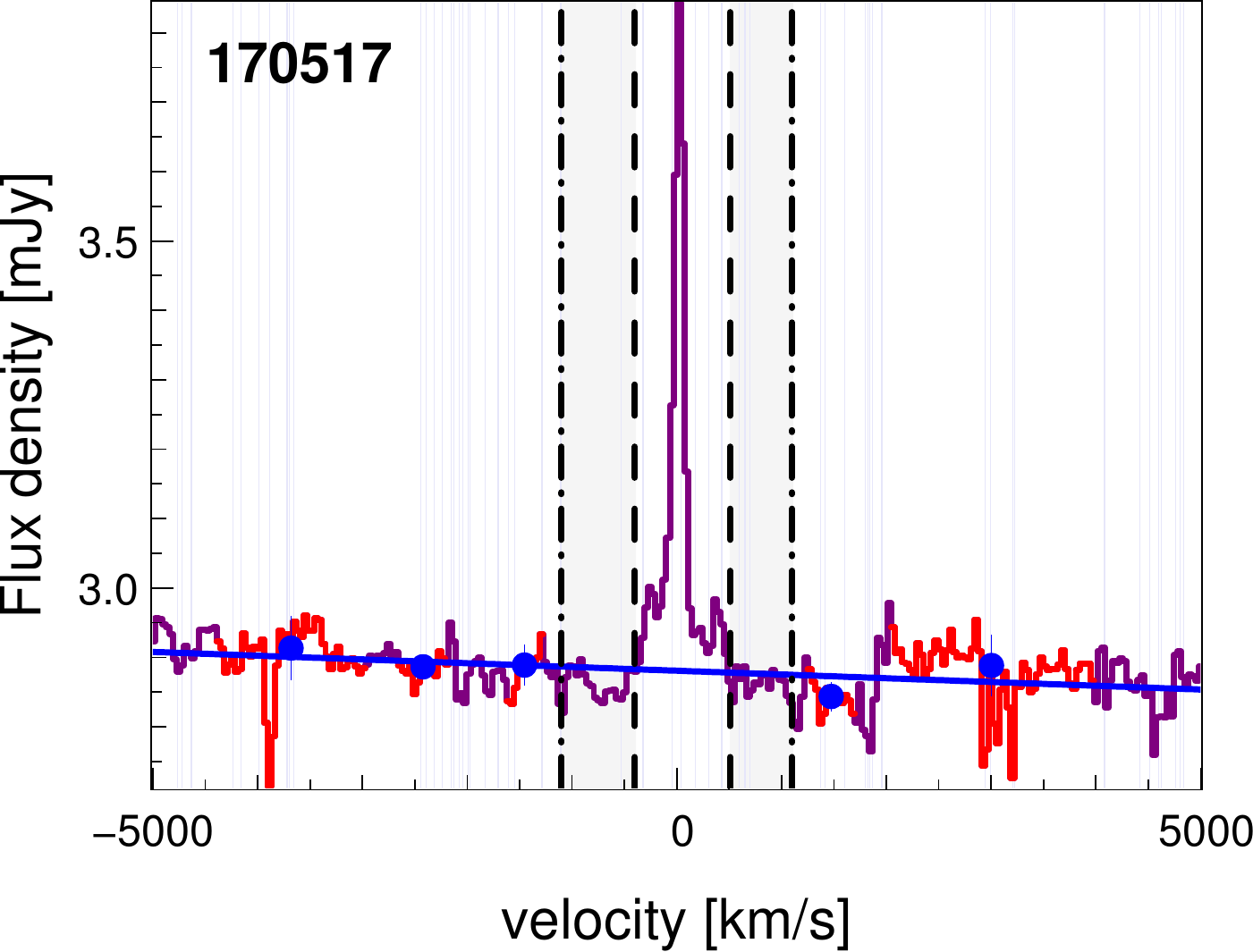}
    \includegraphics[width=4.4cm]{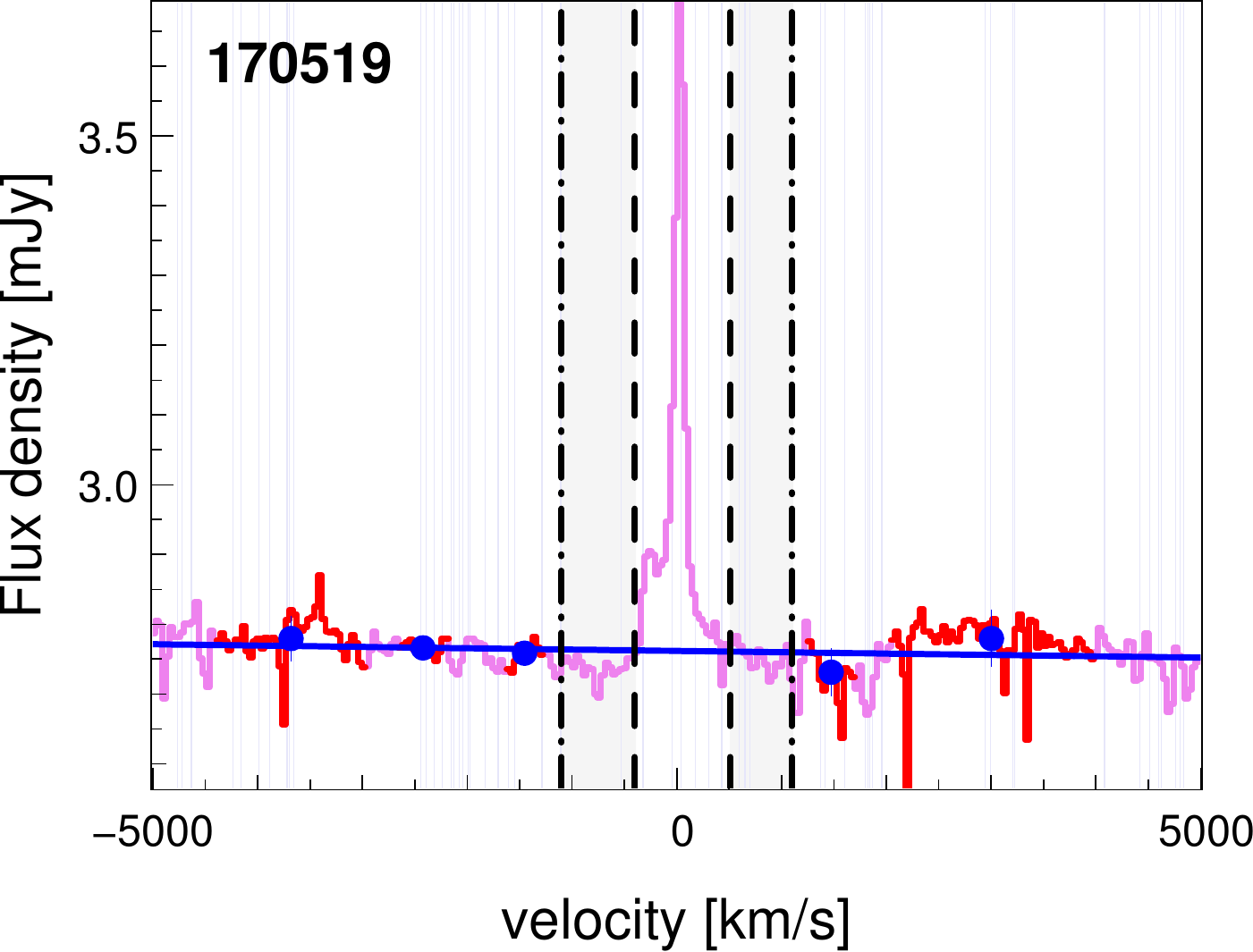}
    \includegraphics[width=4.4cm]{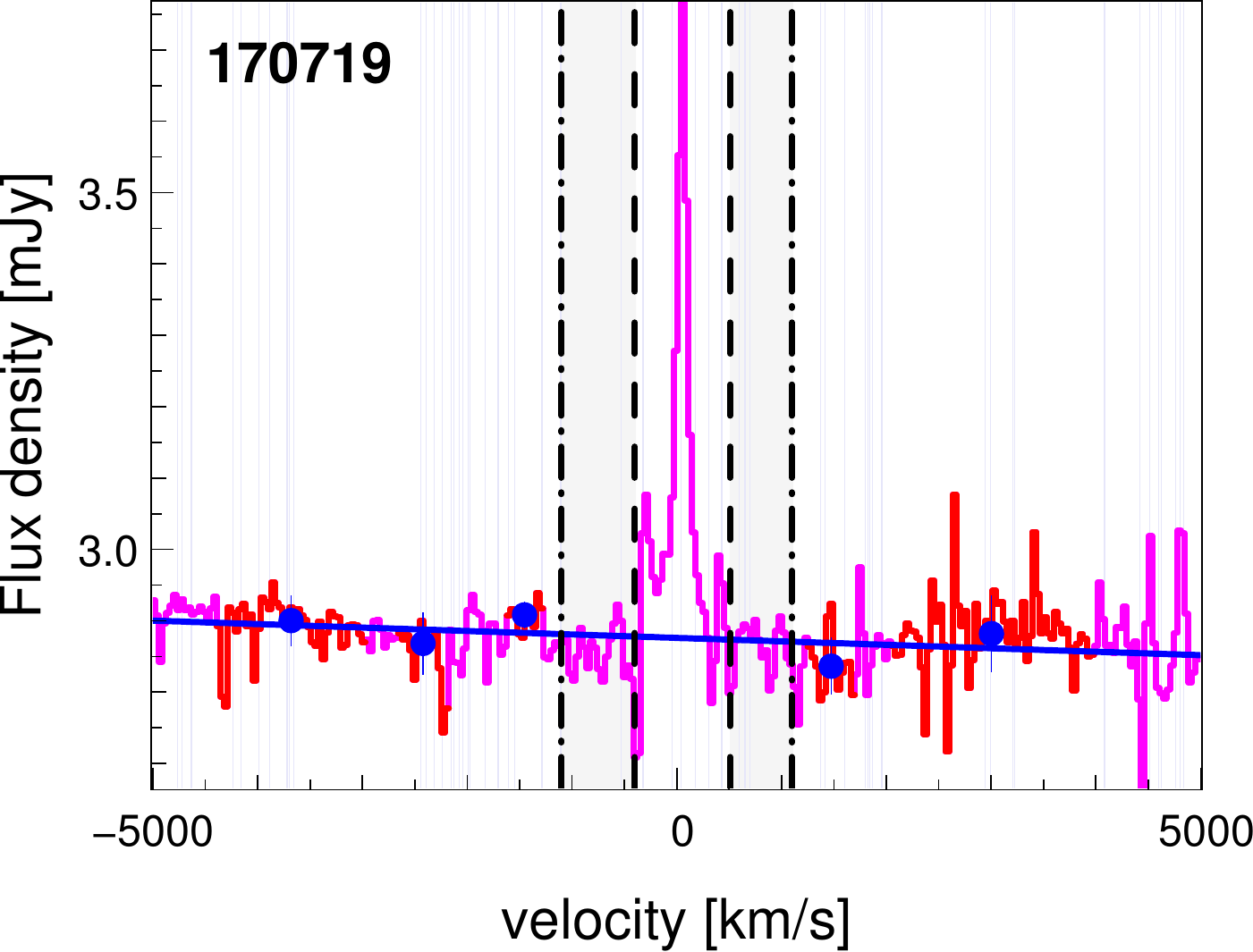}
    \includegraphics[width=4.4cm]{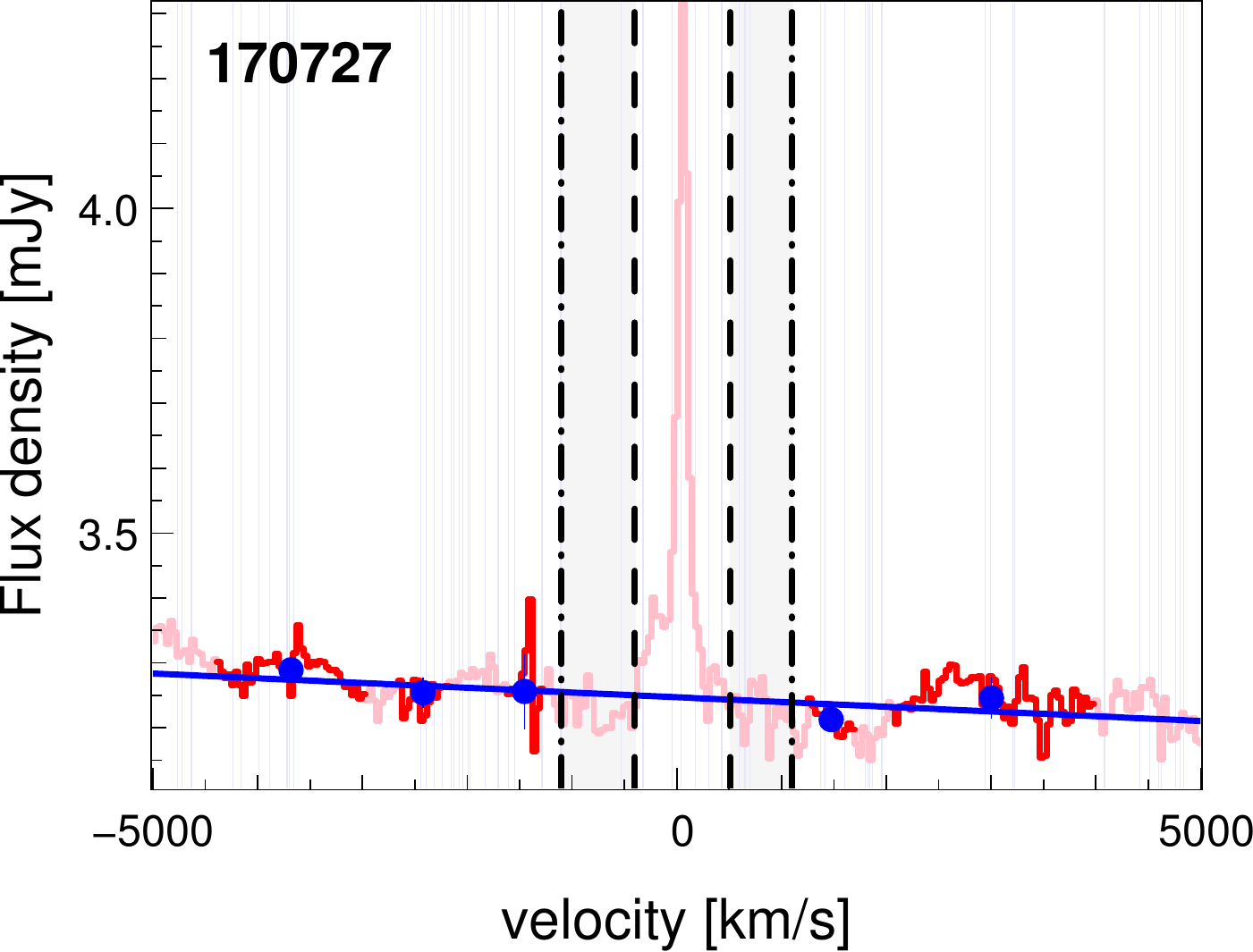}
    \includegraphics[width=4.4cm]{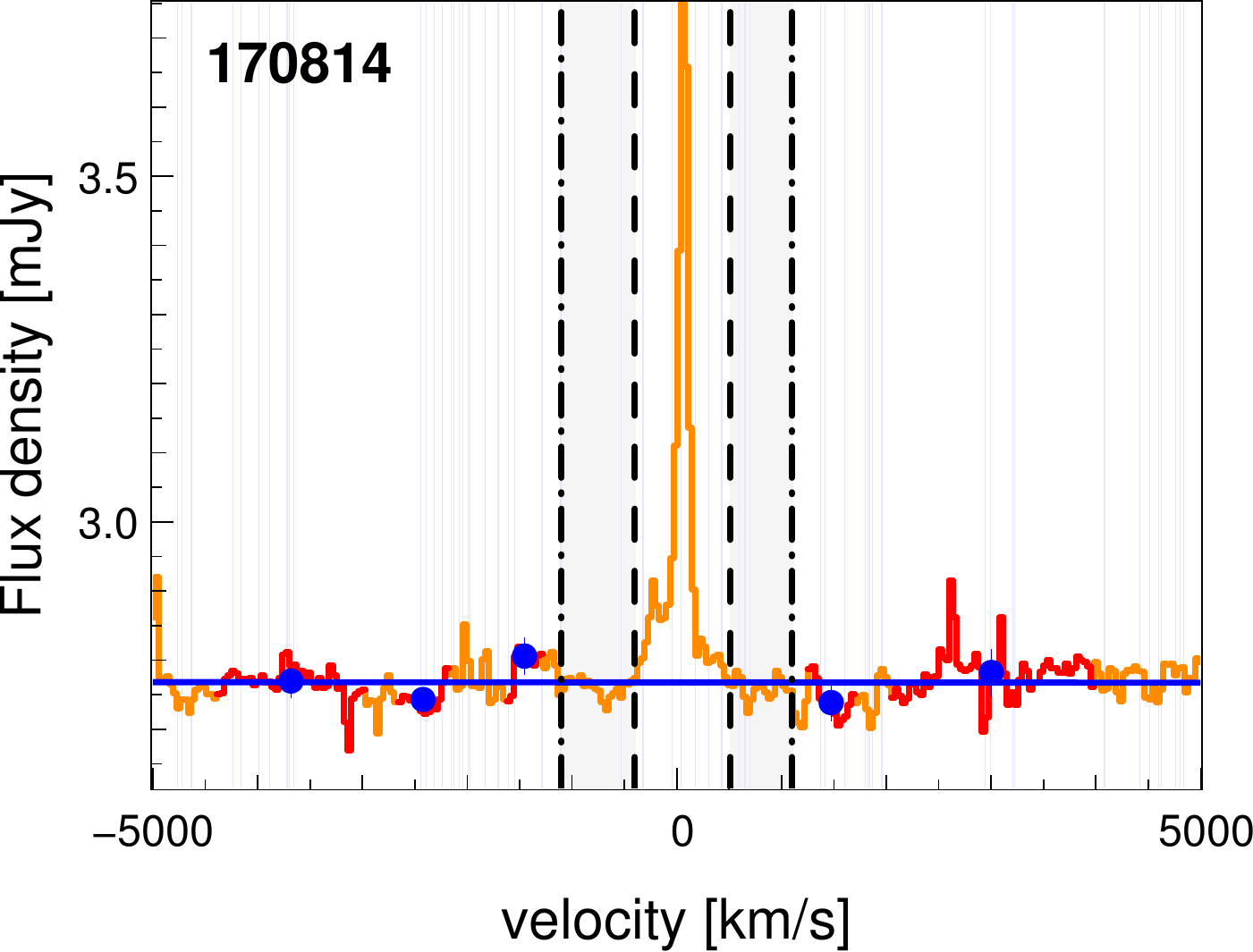}
    \includegraphics[width=4.4cm]{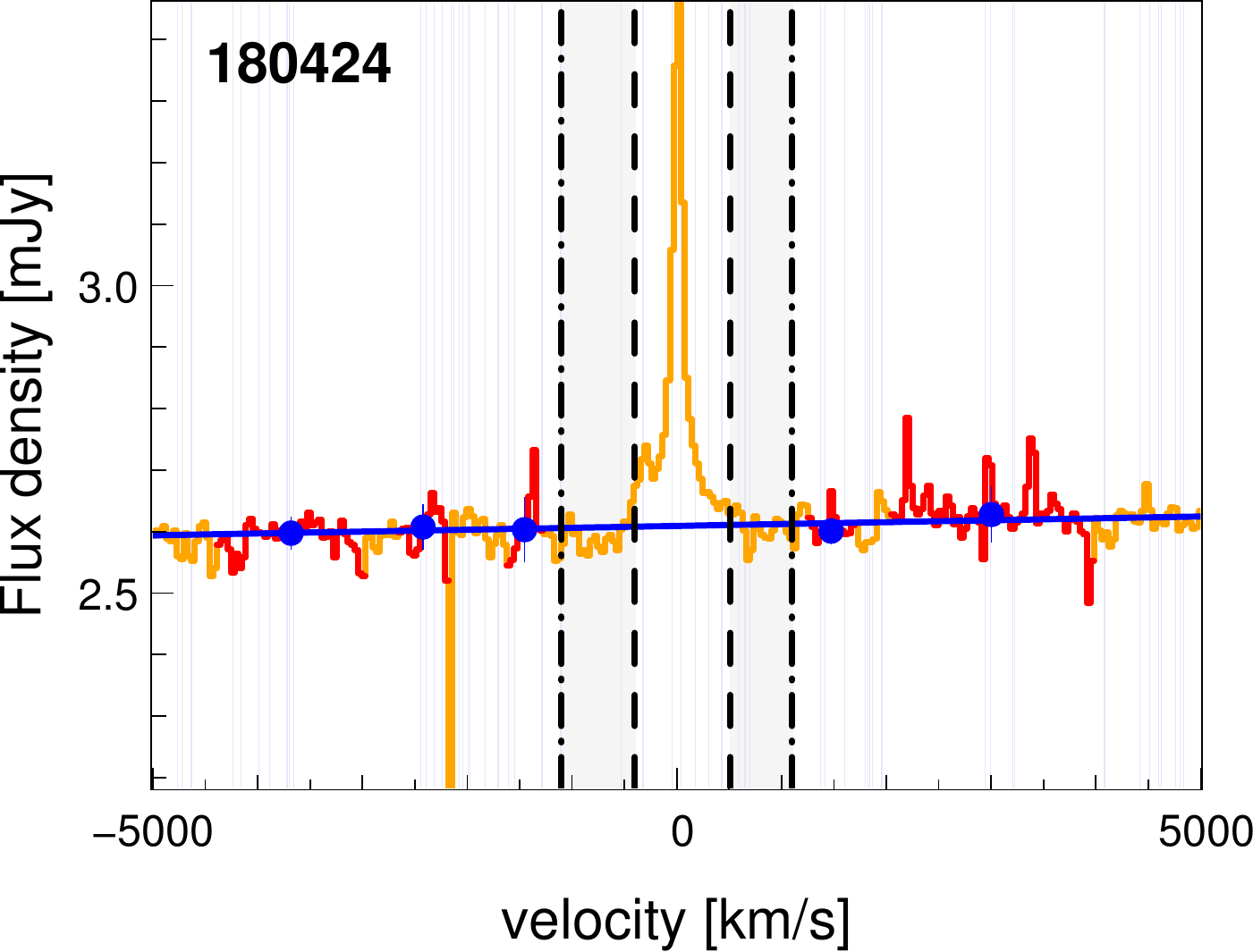}
    \includegraphics[width=4.4cm]{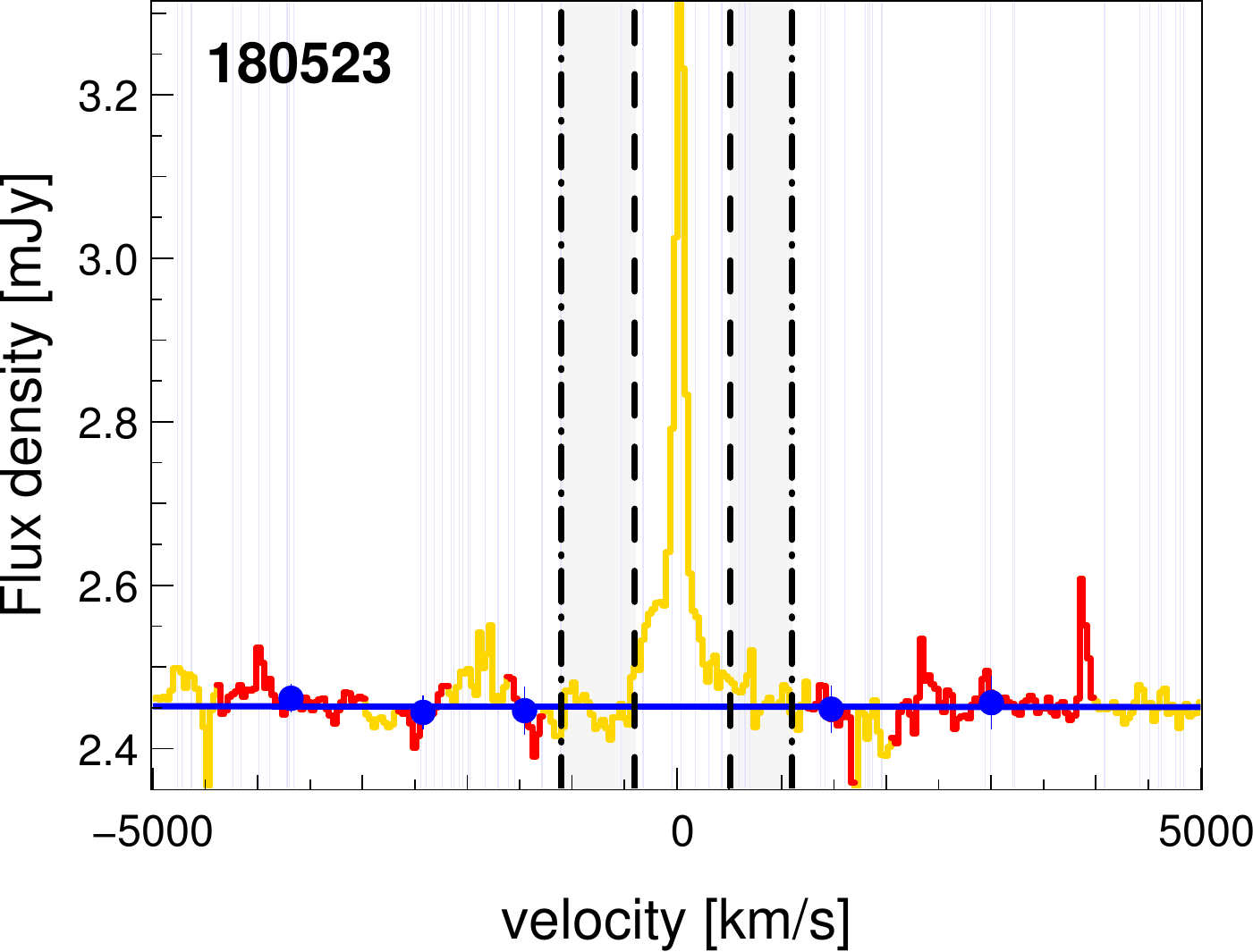}
    \includegraphics[width=4.4cm]{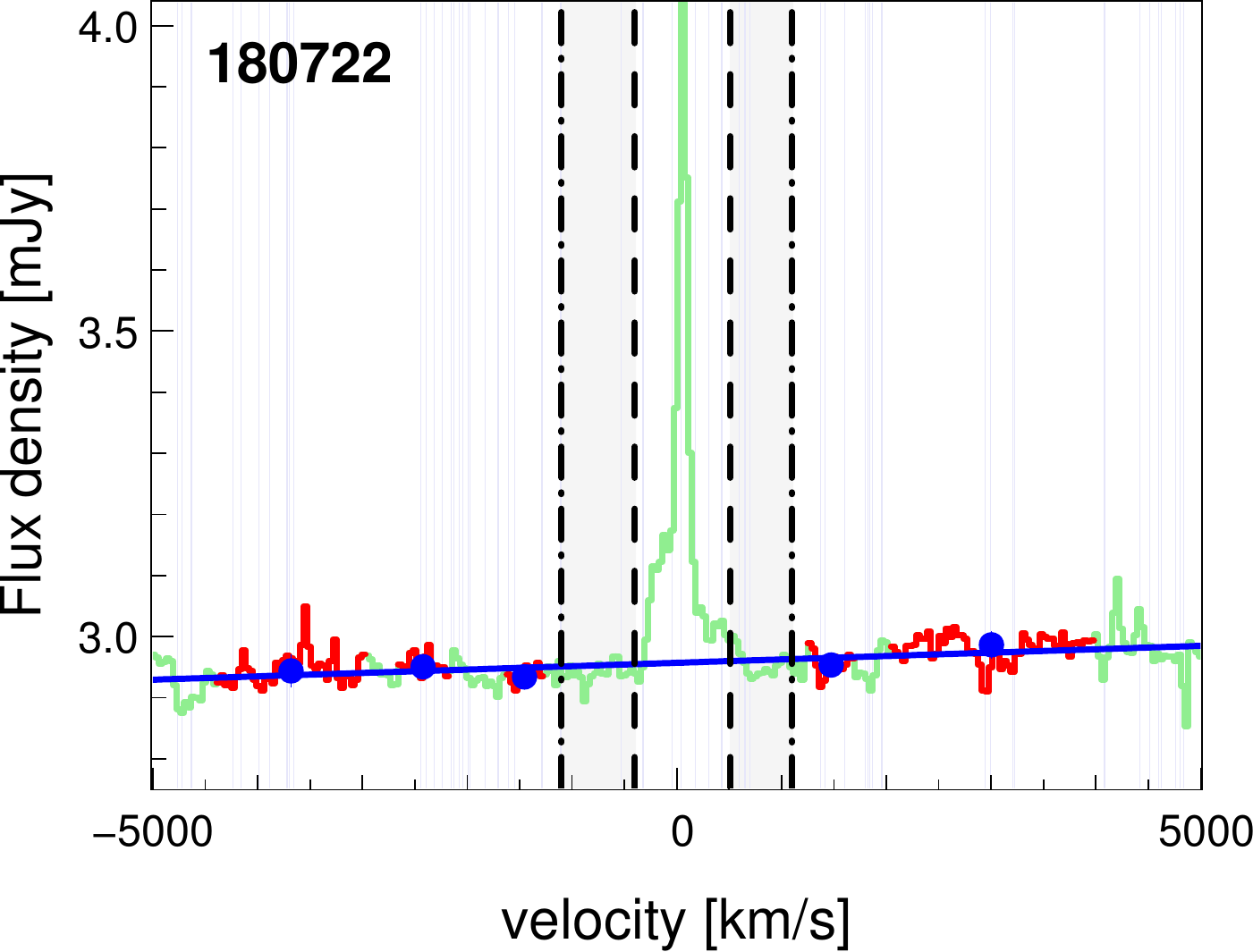}
    \includegraphics[width=4.4cm]{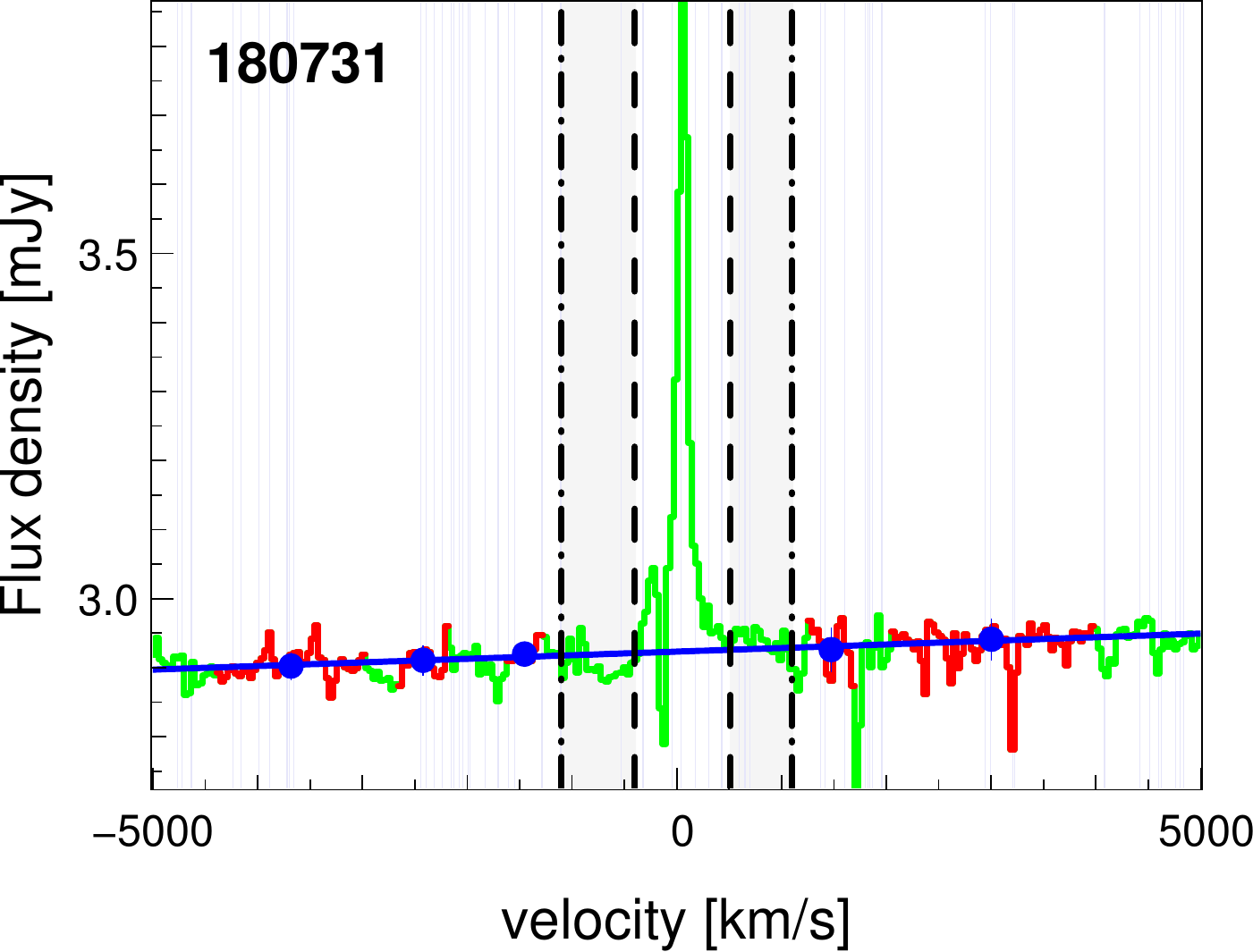}
    \includegraphics[width=4.4cm]{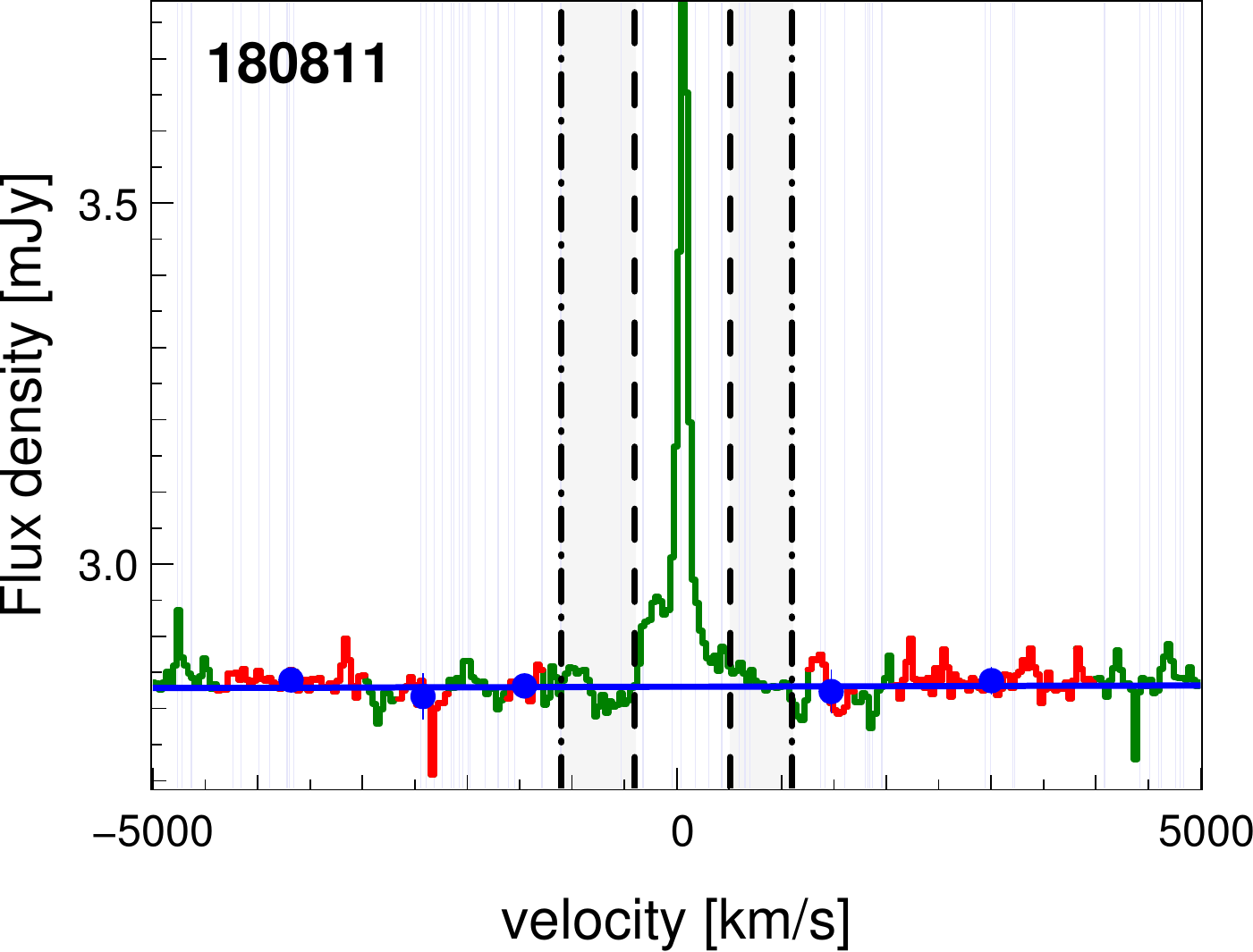}
    \caption{
    Individual spectra and continuum fits. 
    For each epoch (dates reported in YYMMDD format) the spectrum is extracted over a 0.23 '' aperture around Sgr~A*.
    The colors of the individual spectra are the same as in previous figures.
    We consider the median (blue dots) for each of several ranges devoid of spectral features (red) to estimate a linear continuum (blue line). 
    We avoid spectral channels containing strong OH lines (lavender). 
    The dash-dotted vertical lines show the width of the reported H30$\alpha$ line.
    The shaded area represents our notch filter.}
    \label{fig:wholeall}
\end{figure}

In order to isolate the Br$\gamma$ line emission, we removed the stellar continuum from each epoch.
We do that by selecting several wavelength ranges devoid of emission features: from 2.1343 to 2.1447~$\mu$m, from 2.1469 to 2.1503~$\mu$m, from 2.1544 to 2.1568~$\mu$m, from 2.1752 to 2.1783~$\mu$m and from 2.1808 to 2.1948~$\mu$m (displayed in Figure~\ref{fig:wholeall}). 
We then estimate the continuum slope with a linear fit in those ranges, using the same procedure as in \citealt{Ciurlo20}).
In the data reduction each individual exposure is sky-subtracted with a procedure \citep{Davies07} that scales OH lines to be consistent with the science exposures. After that we build the mosaics. There could still be small OH lines residuals after this procedure. 
Because this imperfect sky subtraction could lead to poor continuum estimation, we therefore avoided spectral channels containing bright telluric OH lines within the selected ranges.
The ranges used for the continuum estimation and the continuum fit are shown in Figure~\ref{fig:wholeall}.
The continuum estimation is less reliable near the edges of the Kn3 filter, but that does not affect our analysis, as the broad line we are seeking is close to the center of the spectral band.

To calculate the noise level $\sigma$ we calculate the root-mean-square flux density over ranges devoid of spectral features (the same used for the continuum estimation, see Figure~\ref{fig:wholeall}).
According to the \href{https://www2.keck.hawaii.edu/inst/osiris/idx-preobs.html}{OSIRIS manual}, the average resolution with the 35 mas platescale for the Kn3 band corresponds to $\sim$~2.5 channels.
Therefore, we smooth the spectrum over the width of a resolution element before calculating the root-mean-square noise.
Since the single spectra are extracted over a number of pixels that varies from epoch to epoch (because of the star's mask) we  rescale each individual spectrum to the same effective area. 
In this way the measured noise reflects the intrinsic noise of each spectrum.

\section{Planting H30$\alpha$ profile into the NIR spectrum}
\label{App:plant}
\begin{figure}[htb]
    \centering
    \includegraphics[width=10cm]{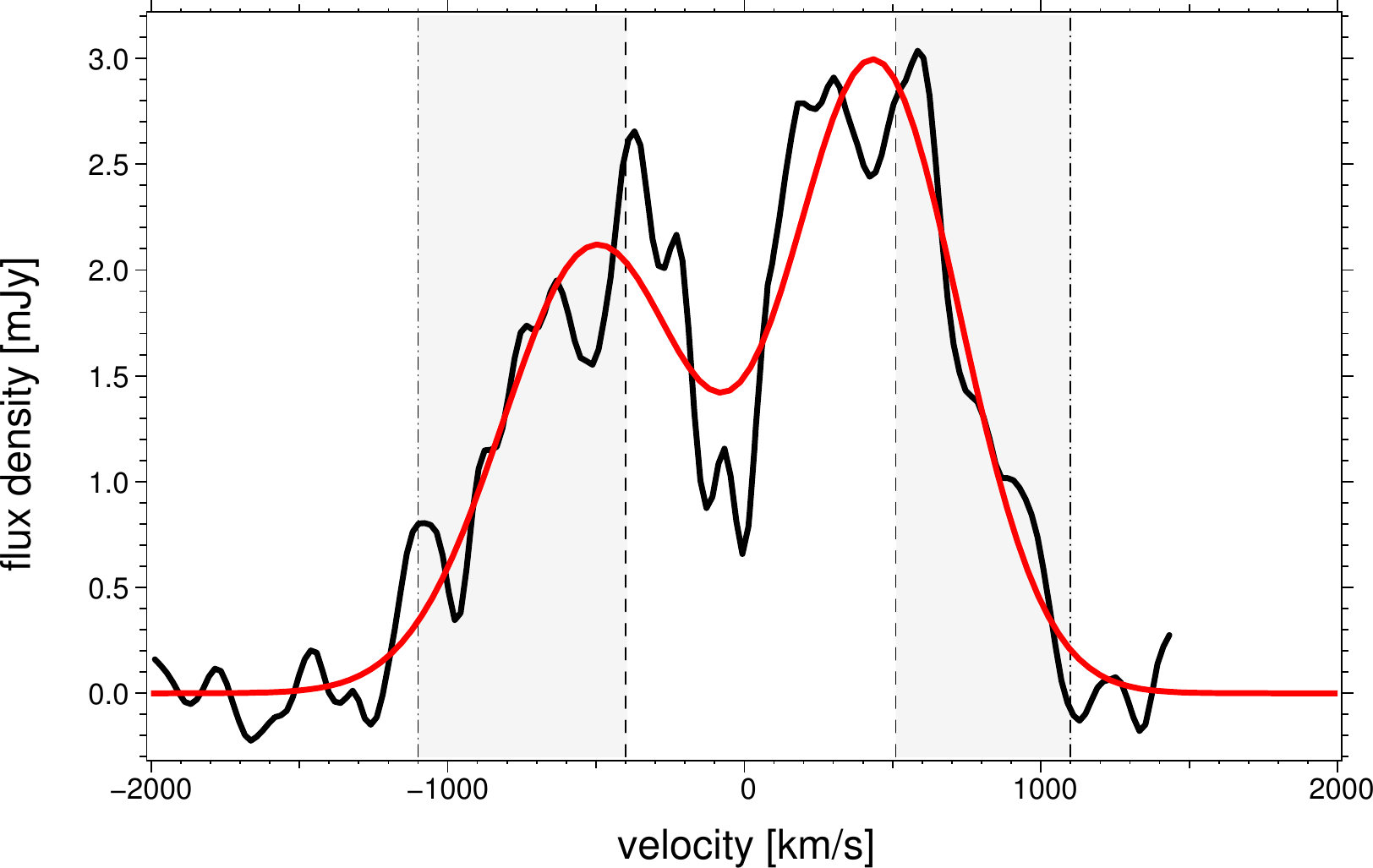}
    \caption{Observed H30$\alpha$ spectrum from \cite{Murchikova19} modeled with a double-peaked Gaussian (red).}
    \label{fig:afit}
\end{figure}
We model the H30$\alpha$ profile with a double-peaked Gaussian (Figure~\ref{fig:afit}). 
To account for the error in fitting the H30$\alpha$ profile with Gaussians, we draw randomly the width and central wavelength within the uncertainties of the fitted values 1000 times.
We vary the flux between 1/100 and 10 times the flux limit estimated from the channel-to-channel noise.  
We calculate the signal-to-noise as the total flux in our notch filter, divided by the square-root of the number of independent resolution elements times the noise per resolution element. 
We identify the total fluxes that fall in the interval of (5$\pm$0.01)$\sigma$ and take the median as our final total flux.
This way we obtain a limit on the Br$\gamma$ flux of 1.13~10$^{-19}$~W/m$^2$ in our notch filter.
We estimate the uncertainty as the root-mean square of the values one obtains by running the procedure 19 times, dropping at each time one of the 19 epochs of data.

\section{The extended gas emission}
\label{App:gas}
\begin{figure}
    \centering
    \includegraphics[width=10cm]{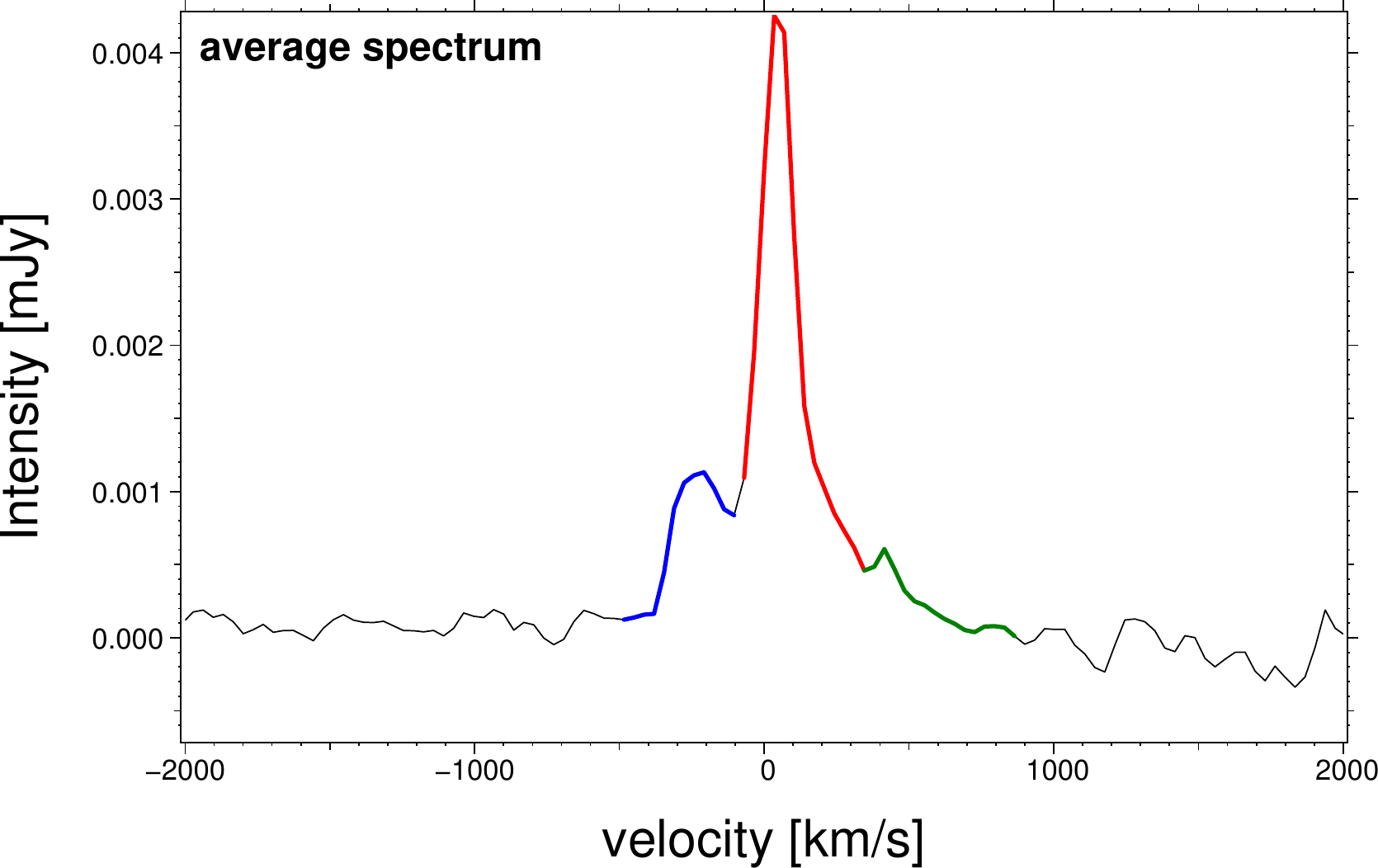} \\
    \vspace{.5cm}
    \includegraphics[width=5.cm]{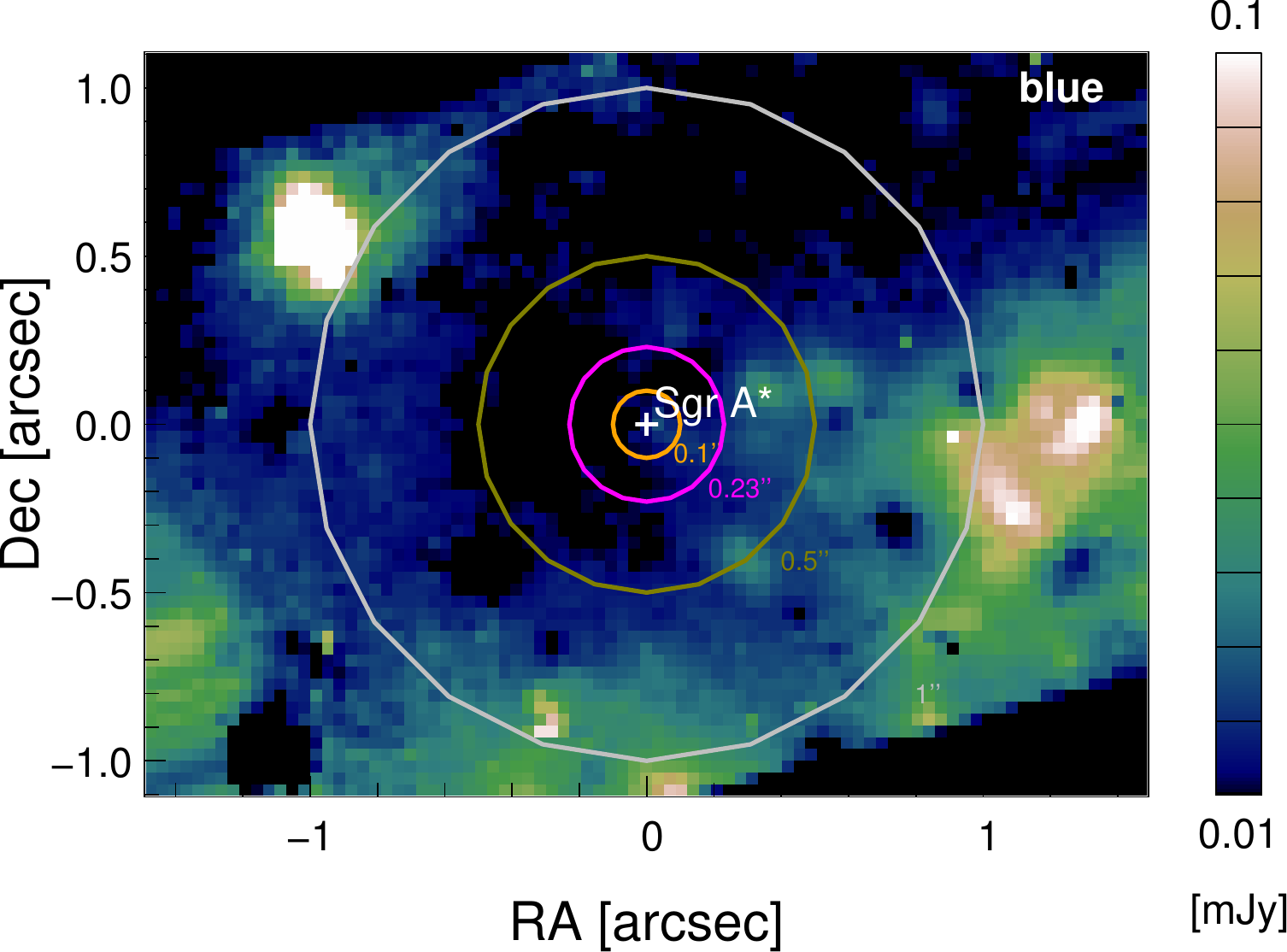}
    \includegraphics[width=5.cm]{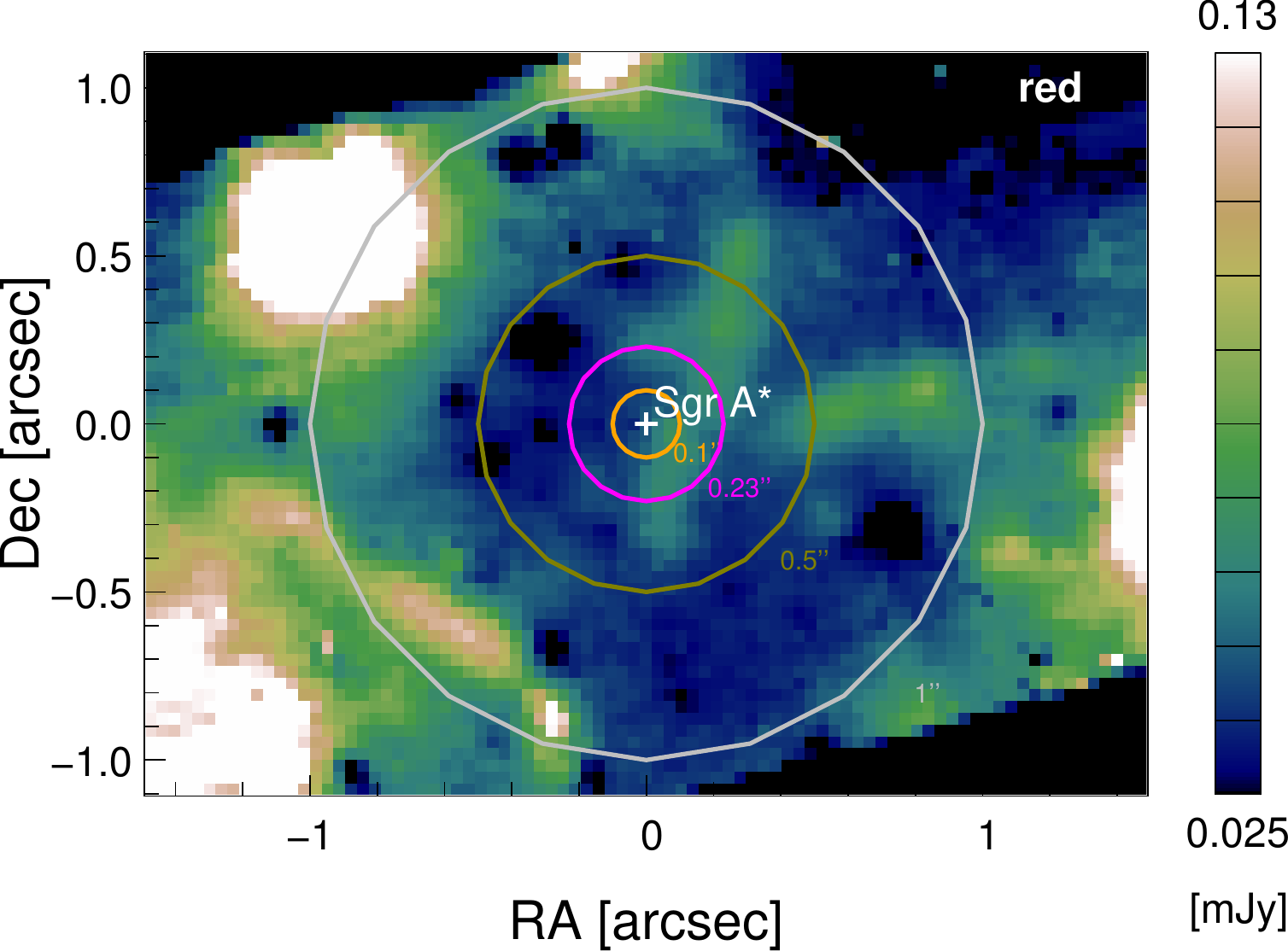}
    \includegraphics[width=5.cm]{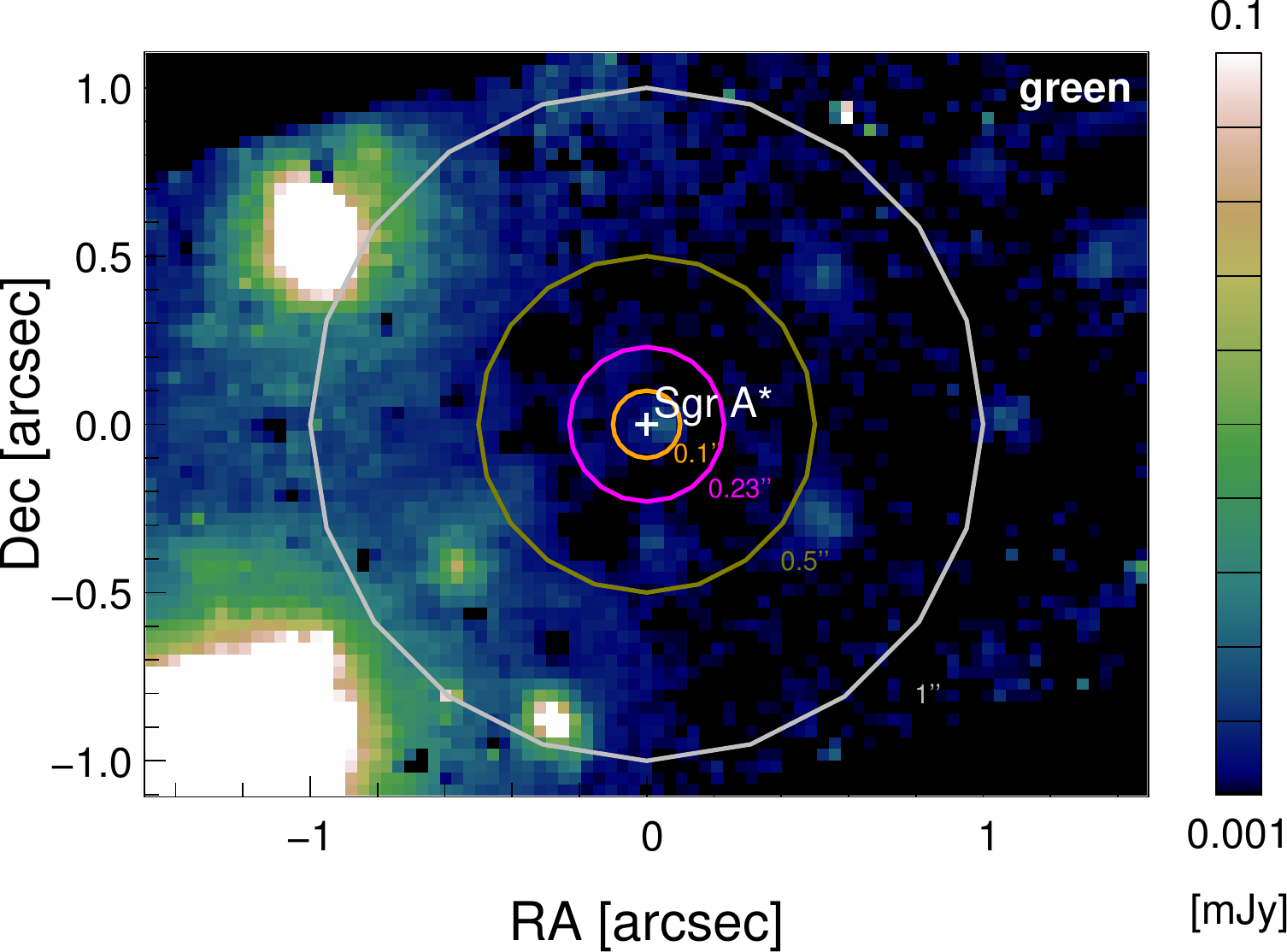} \\
    \vspace{.5cm}
    \includegraphics[width=9cm]{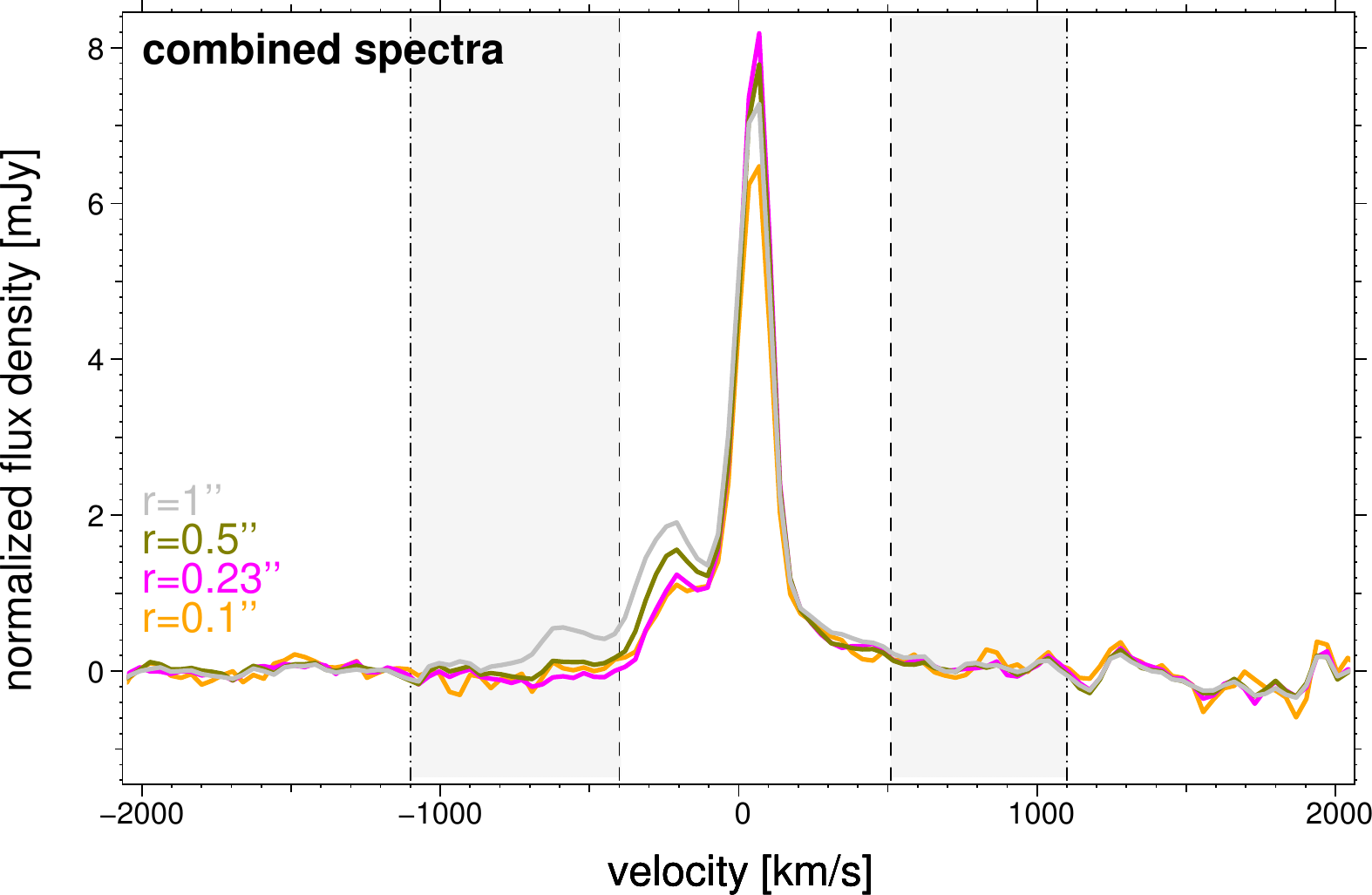}
    \includegraphics[width=8cm]{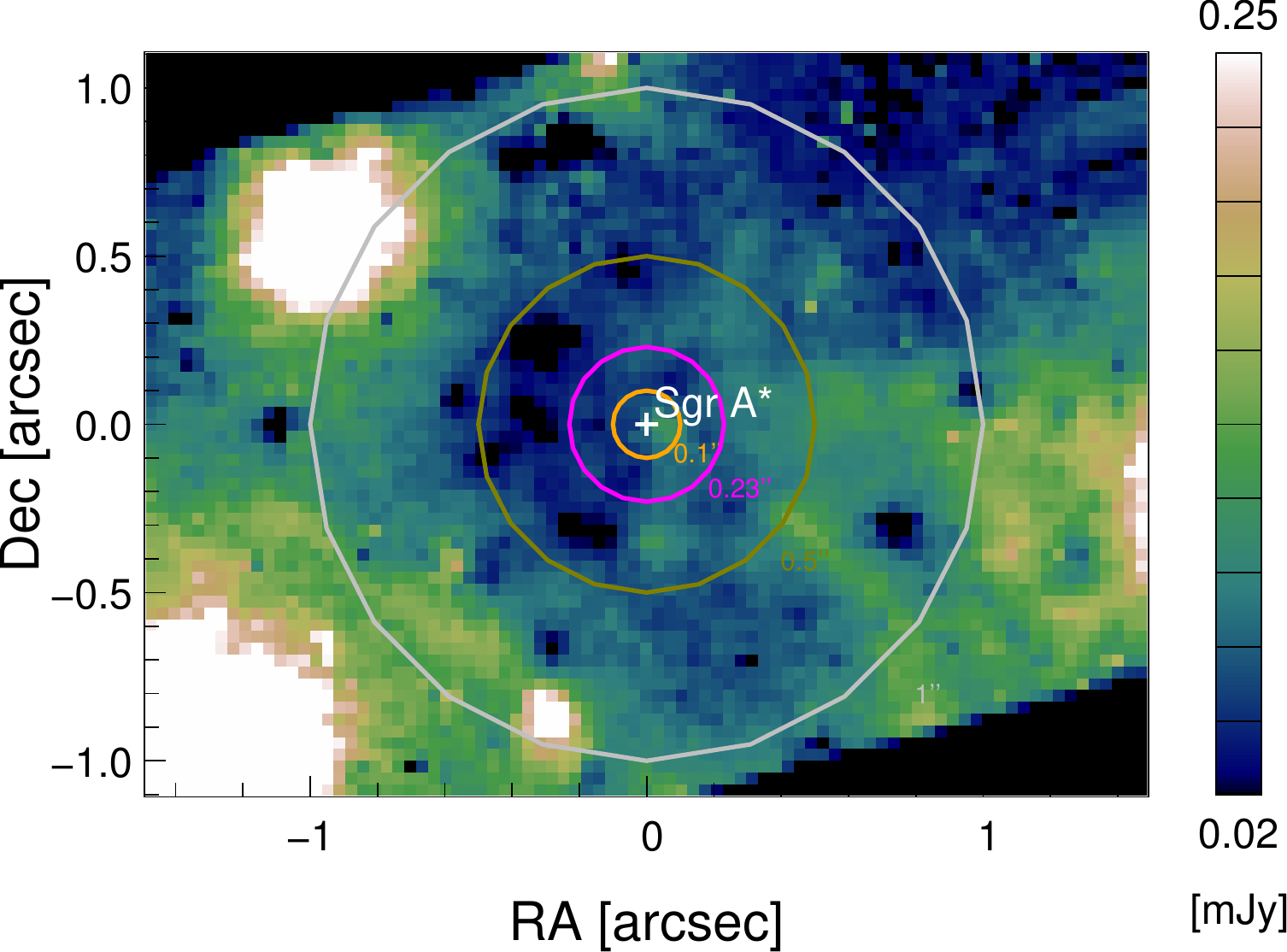}
    \caption{
    Top: Average spectrum extracted from the 2017 continuum-subtracted cube within a 1'' aperture radius. 
    The different gas components are highlighted in blue, red and green. 
    Middle: gas distribution associated with the different gas components (2017 cube collapsed over the spectral ranges corresponding to each gas component). 
    Bottom left: combined continuum-subtracted spectra extracted over several radii.
    The shaded velocity ranges represent our notch filter.
    Each combined spectrum shows the total flux enclosed in the aperture, normalized by the enclosed area in the aperture.
    Bottom right: 2017 cube collapsed over our notch filter. 
    }
    \label{fig:gas}
\end{figure}

The OSIRIS data show that there are multiple extended foreground and background gas components to the Br$\gamma$ emission in this region (evident in Figure~\ref{fig:wholeallr}, near the Br$\gamma$ rest frequency).
The gas observed toward SgrA~* has complex, multi-component emission extended in space and velocity.
We can identify at least three components (see Figure~\ref{fig:gas} top and middle): one associated with nearby radio source Epsilon \citep{Yusef-Zadeh90}, one associated with the bright Wolf-Rayet stars in the field (IRS16~C and IRS16~SW) and one featuring an almost linear feature immediately adjacent to the position of Sgr~A* along the line of sight \citep{Schodel11, Peissker20}. 
Some of these components move enough from year to year to require a specific fit for each epoch if they are to be subtracted from the spectrum.
However, given that our goal is to measure or limit very broad Br$\gamma$ emission ($\sim$2000~km/s) and given that the foreground and background emission occurs at much lower velocities ($<$400--500~km/s),  
we simply exclude the low-velocity range (between -400 and +510~km/s) from our analysis of the Br$\gamma$ emission.  
We note that by changing the bright star mask from epoch to epoch we sample different portions of the confusing background in different epochs. 
However, the wavelength extent of the gas does not exceed the boundary of the exclusion zone we set and therefore it does not influence our analysis.
Figure~\ref{fig:gas} (bottom) shows how, at larger radii, this superimposed gas emission becomes more and more challenging to deal with.
Therefore, for this work we only focus on a 0.23'' aperture that allows us compare our limits directly to the \cite{Murchikova19} observations.

\end{document}